\documentclass[12pt]{article}
\usepackage{amssymb,amsfonts,amsthm,dsfont,mathtools}
\usepackage{amsmath}
\usepackage{booktabs,caption,fixltx2e}
\usepackage[flushleft]{threeparttable}
\usepackage{multirow}
\usepackage{setspace}
\usepackage{color}
\usepackage{endnotes}
\usepackage{enumerate}
\usepackage{float}
\usepackage[bottom]{footmisc}
\usepackage[pdftex]{graphicx}
\usepackage{epstopdf}
\usepackage{tikz}
\usepackage{indentfirst}
\usepackage{latexsym}
\usepackage{lscape}
\usepackage{parskip}
\usepackage{rotating}
\usepackage{caption}
\usepackage{subfigure}
\usepackage{bm}
\usepackage{verbatim}
\usepackage{mathdots}
\usepackage[authoryear, round,semicolon,sort&compress]{natbib}
\usepackage{soul}
\usepackage[left=0.75in,right=0.75in,top=0.75in,bottom=0.75in]{geometry}
\usepackage{multirow}
\usepackage{mwe}
\usepackage{arydshln}
\usepackage[scaled=0.85]{beramono}
\usepackage{datetime}

\allowdisplaybreaks[4]

\newtheorem{dfn}{Definition}[section]
\newtheorem{asn}{Assumption}[section]
\newtheorem{lma}{Lemma}[section]
\newtheorem{thm}{Theorem}[section]

\newtheorem{cor}{Corollary}[section]

\makeatletter
\newcommand{\leqnomode}{\tagsleft@true}
\newcommand{\reqnomode}{\tagsleft@false}
\makeatother

\newcommand\T{\rule{0pt}{2.5ex}}

\newcommand{\var}{\text{Var}}
\newcommand{\argmin}{\mathop{\text{argmin}}}
\newcommand{\argmax}{\mathop{\text{argmax}}}
\newcommand{\cov}{\text{Cov}}

\newcommand{\eps}{\varepsilon}

\allowdisplaybreaks[2]
\usepackage[colorlinks=true,allcolors=blue]{hyperref}
\makeatletter
\def\mathcenterto#1#2{\mathclap{\phantom{#1}\mathclap{#2}}\phantom{#1}}
\let\old@widetilde\widetilde
\def\widetildeto#1#2{\mathcenterto{#2}{\old@widetilde{\mathcenterto{#1}{#2\,}}}}
\let\old@widehat\widehat
\def\widehatto#1#2{\mathcenterto{#2}{\old@widehat{\mathcenterto{#1}{#2\,}}}}
\makeatother

\def\widetilde{\widetildeto{a}}

\begin{document}

\sloppy

\title{Identification of Linear Regressions with Errors in all Variables}
\author{Dan Ben-Moshe\thanks{{\small Email: danbm@huji.ac.il.	
			The paper is loosely based on 
			the second chapter of my PhD thesis. 
			I am grateful to Rosa Matzkin and Jinyong Hahn for their generous support, advice, and guidance.
			I thank the editor Liangjun Su as well as Xavier D'Haultf\oe{}uille, Jinyong Hahn, Allyn Jackson, Arthur Lewbel, Rosa Matzkin, Yannay Spitzer, and Elie Tamer for reading various parts of drafts and providing feedback.
			I benefited from comments at Aarhus, Boston College, CREST, Harvard, Hebrew University of Jerusalem, and Penn State, and
			discussions with  David Genesove, Patrik Guggenberger, Kei Hirano, Stefan Hoderlein, Max Kasy, Saul Lach,  \'{A}ureo de Paula, 
			Anna Simoni, James Stock, Daniel Wilhelm, and Martin Weidner.
			} }\\
{\small The Hebrew University of Jerusalem}}
\date{March 2020}
\maketitle

\begin{abstract}
 
This paper analyzes the classical linear regression model with measurement errors in all the variables. First, we provide necessary and sufficient conditions for identification of the coefficients. We show that the coefficients are not identified if and only if an independent normally distributed linear combination of regressors can be transferred from the regressors to the errors. Second, we introduce a new estimator for the coefficients using a continuum of moments that are based on second derivatives of the log characteristic function of the observables. In Monte Carlo simulations, the estimator performs well and is robust to the amount of measurement error and number of mismeasured regressors. In an application to firm investment decisions, the estimates are similar to those produced by a generalized method of moments estimator based on third to fifth moments.

\end{abstract}


\pagebreak

\section{Introduction}\label{se:intro}

The efficacy of the classical linear regression model, which stands at the heart of applied economics, can be compromised by measurement errors. 
Despite their pervasiveness in data and the resulting endogeneity problems, measurement errors are mostly ignored in practice.

The linear errors-in-variables model generalizes the linear regression model in order to take measurement errors into account.\footnote{See \cite{aigner1984latent}, \cite{buonaccorsi2010measurement}, \cite{carroll2006measurement}, \cite{chen2011nonlinear}, \cite{cheng1999statistical}, \cite{fuller2009measurement}, \cite{gillard2010overview}, and \cite{Schennach2016review} for reviews of measurement error models.}  For the single regressor errors-in-variables model, the conditions under which the coefficient can be identified and estimated are well understood.
Surprisingly, these conditions are not known for the multiple regressor errors-in-variables model.  This is the problem the present paper addresses. 

In the case of the single regressor errors-in-variables model, \cite{reiersol1950} shows that the coefficient is not identified if and only if (i) the regressor is normal and (ii) either one of the errors can be expressed as the sum of two independent variables, one of which is normal. 
Hence, identification requires some nongaussianity assumption and consistent estimation is usually based on higher moments. 
Accurate estimation of the coefficient based on higher moments, however,  requires large samples and is particularly sensitive to outliers.  
Common practice is to ignore higher moments and use the standard ordinary least squares estimator, based on second moments, 
which bounds the unknown coefficient and identifies its sign.

The general conditions for identification of the multiple regressor errors-in-variables model are not known. 
The main contribution of this paper is to provide necessary and sufficient conditions for identification of the coefficients. 
The coefficients are not identified by first and second moments so that jointly normal observables do not identify the coefficients \cite[see, e.g.,][]{koopmans1937linear,reiersol1945confluence}. 
Further, unlike the single regressor model,
first and second moments do not in general bound the coefficients, 
unless they all lie in the same orthant \citep[see][]{klepper1984consistent}.
The literature  
has based identification on nongaussianity assumptions on the unobserved regressors. 
 \cite{cragg1997using}, \cite{Dagenais97}, \cite{erickson2002two},
  \cite{geary1941inherent}, \cite{lewbel2012using}, and \cite{pal1980consistent} identify the coefficients using rank conditions on third and higher moments.
\cite{bekker1986comment}, \cite{kapteyn1983identification}, and \cite{willassen1979extension}
identify the coefficients using the assumption that no linear combinations of the regressors are normal. 

We show that the model may still be identified even if the unobserved regressors are jointly normal. 
For nonidentification, a linear combination of regressors must be not only normal but also transferable to the errors.
Specifically, we show that the coefficients are not identified if and only if (i) there is a normally distributed linear combination of unobserved regressors that is (ii) independent of some other linear combinations of regressors and (iii) some errors can be expressed as the sum of two independent random variables, one of which is normal.

Our identification strategy uses characteristic functions to take advantage of the model's linearity in the regressors and errors. 
The literature that identifies measurement error models that are nonlinear in the regressors uses additional information such as instruments or repeated measurements \citep[see, e.g,][]{hausman1991,hu2015closed,Schennach07}. 
A notable exception is \cite{SchennachHu13} who show identification of a nonparametric function with a scalar mismeasured regressor.  
Extending identification to nonlinear functions of multidimensional mismeasured regressors, without any additional information, would be a new result.

The second contribution of this paper is to introduce a new estimator for the coefficients that uses information from all higher moments and that is not based on a solution to a system of higher moment equations which, as mentioned earlier, requires large samples for accurate estimation.\footnote{In addition to estimation methods based on higher moments, estimation can be based on, for example, distributional assumptions, instrumental variables, multiple measurements, small error variance, etc.  \citep[e.g.][]{bickel1987efficient,spiegelman1979estimating,hausman1991,li2002robust,schennach2004estimation,chesher1991effect}.} 
Although identification is not constructive (i.e., identification does not produce a closed form mathematical expression for the unknown coefficients), we will nevertheless be able to
mimic the steps in the identification procedure to construct an estimator using a continuum of moment equations based on second order partial derivatives of the log characteristic function of the observables. 
The estimator is root-$n$ consistent and asymptotically normal. 
 
The identification and estimation results allow all the regressors to have measurement errors. 
If there are some regressors that are known to be perfectly measured 
then there are other ways to identify the coefficients. For example, \cite{lewbel1997constructing} constructs instruments using the perfectly measured regressors and \cite{benmoshe2017meas} project off the perfectly measured regressors and use them as instruments. 
The simulations and application focus on the more empirically relevant setting in which some of the regressors are mismeasured but others are not. 
We estimate the coefficients by first partialling out the perfectly measured regressors and then applying our results to the residuals of regressions of the outcome variable and mismeasured regressors on the perfectly measured regressors. 
In the finite sample simulations the estimator performs well with small bias and decreasing standard deviation at root-$n$ rate. The coverage probabilities of the 95\% confidence interval are close to the nominal level suggesting that inference using the estimator is trustworthy. 
The estimates are robust to the amount of measurement error, the number of mismeasured regressors, and the distribution of the regressors.

Using data on firm investment decisions, the estimator produces results that are similar to the generalized method of moments estimator based on third, fourth, and fifth moments.
According to the 
neoclassical theory of firm investment, 
these decisions should depend only on marginal returns to capital \citep{tobin1969general,kaldor1966marginal,brainard1968pitfalls,hayashi1982tobin,mussa1977external,abel1983optimal,abel1994uni}.
Empirical studies, however, show that measures of marginal returns to capital explain very little about firm investment decisions \citep[see, e.g.,][]{blanchard1994firms}, and have spurred
a long-standing debate on why the theory is not confirmed by the data.
\citet{erickson2000measurement,erickson2014minimum} provide evidence that a linear regression with
measurement error in marginal returns to capital corroborates the theory.
However, most of the literature contends that financial constraints need to be incorporated into the theory \citep[see, e.g.,][]{gilchrist1995evidence,fazzari1988investment}.
These financial constraints are unobservable and often measured by cash flows. 
We show that
the estimated coefficients are mostly insignificant in a linear regression with measurement errors in both marginal returns to capital 
and in financial constraints. 
This suggests that either the theory is missing something important or the linear errors-in-variables model is inappropriate.
 
The paper is organized as follows. Section \ref{se:eivid} contains the assumptions and identification results.
Section \ref{se:eivest} presents a root-$n$ consistent and asymptotically normal estimator for the coefficients.
In finite sample simulations in Section \ref{se:eivsim} and in data on firm investment decisions in Section \ref{se:app}, 
we compare our integrated generalized method of moments (IGMM) estimator to a generalized method of moments (GMM) estimator based on third to fifth moments and the ordinary least squares (OLS) estimator. 
All the proofs are contained in the appendix.

\textbf{Notation.} The main tool of analysis is the log characteristic function $\varphi_{\bm X^*}(\bm u)=\ln (E[e^{i\bm u' \bm X^*}])$, where
$\varphi_{\bm X^*}(\bm u)$ completely defines the $p$-dimensional random vector $\bm X^*$,
$\bm u'$ is the transpose of $\bm u \in \mathbb{R}^{p}$, ${i}=\sqrt{-1}$ is the imaginary unit, $\ln(\cdot)$ is the natural log restricted to its principal branch, and $E[\cdot]$ is the expectation operator. 
The log characteristic function of 
a jointly normal random vector $\bm X^* \sim N(\bm \mu_{\bm X^*},\bm \Sigma_{\bm X^*})$ with mean $\bm \mu_{\bm X^*}$ and covariance matrix 
$\bm \Sigma_{\bm X^*}$ 
is $\varphi_{\bm X^*}(\bm u)=i\bm \mu_{\bm X^*}'\bm u -\frac{1}{2}  \bm u'\bm \Sigma_{\bm X^*} \bm u$.
We use the convention that $\bm \Sigma_{\bm X^*}$ can be zero (i.e., a normal random vector can be degenerate, unless otherwise stated).
The matrix $\bm \Sigma_{\bm X^*}$ is positive definite if $\bm u'\bm \Sigma_{\bm X^*}\bm u>0$ for all $\bm u \neq \bm 0$ and
we sometimes write this as $\bm \Sigma_{\bm X^*} \succ 0$.
The log characteristic function is sometimes a polynomial in $\bm u$ of degree $k$ denoted by $P_k(\bm u)$, where $P_0$ is some generic constant.

Matrices and vectors are bolded. 
The concatenation of the matrices $\bm A_1$ and $\bm A_2$ is denoted by $(\bm A_1  ,  \bm A_2)$. 
The $p \times p$ identity matrix is denoted by $\bm I_p$, the standard basis vector that has a $1$ in the $t$-th coordinate and $0$'s elsewhere is denoted by $\bm e_t =(0,\ldots,0,1,0,\ldots,0)'$, 
and a matrix of $0$'s is denoted by $\bm 0$. The dimensions of $\bm e_t$ and $\bm 0$ are the ones needed to make sense of an expression. The Euclidean norm of a vector $\bm u$ is $||\bm u||=\sqrt{\bm u'\bm u}$ and the operator norm of a matrix $\bm A$ is 
$||\bm A|| =\sup_{||\bm u||=1} ||\bm A \bm u ||$. 

\section{Identification}  \label{se:eivid}

We provide necessary and sufficient conditions for identification of the coefficients
in the classical linear errors-in-variables model, without any additional information such as instrumental variables or repeated measurements,
\begin{align}
\left\{\begin{array}{rcl}
\bm X &=& \bm X^* + \bm \eta, \\
Y &=& \bm \beta' \bm X^* + \eps ,
\end{array} \right.  \label{eq:main} 
\end{align}
where $\bm X$, $\bm X^*$, and $\bm\eta$ are real $p$-dimensional random vectors and $Y$ and $\eps$ are real scalar random variables. 
The observable outcome $Y$ is a linear combination of the unobservable regressors $\bm X^*$ and the unobservable structural error $\eps$
while $\bm X$ contains observable measurements of $\bm X^*$ 
contaminated with unobservable measurement errors $\bm\eta$.
The vector $\bm \beta \in \mathbb{R}^p$ contains unknown nonzero coefficients of interest.
We assume that there does not exist a nonzero 
$\bm b \in \mathbb{R}^p$ such that $\bm b'\bm X^*$ 
is degenerate; otherwise there is multicollinearity and the model is not identified even without measurement errors.

We now introduce the distributional assumptions. 

\medskip
\begin{asn}\label{as:eivdep} Let $\bm X^*$ and $(\bm \eta,\eps)$ be independent  and either let
	\begin{enumerate}[(i)]
		\item 
			$\eta_1,\ldots,\eta_p$, and $\eps$ be mutually independent or
			\item 
			$(\bm \eta,\eps)\sim N(\bm 0,\bm \Sigma)$	
			with positive definite covariance matrix $\bm \Sigma$. 
	\end{enumerate}
	
\end{asn}
\medskip

The assumption states that the regressors $\bm X^*$ and the errors $(\bm \eta,\eps)$ are independent 
and that either the errors are mutually independent or the errors are jointly normal with arbitrary covariance matrix. 
If the regressors and any of the errors are allowed to be dependent then the coefficients are not identified without additional information.
If we assume that the regressors and the errors are independent but impose no further assumptions on the errors then the log characteristic function of the errors satisfies a partial differential equation with many functional solutions.
The techniques in this paper identify the coefficients when the log characteristic function of the errors takes the form
$\varphi_{\bm \eta,\eps}(\bm s)=P_{a}(\bm s) + \sum_{k=1}^{p+1}h_k(s_{k})$, $\bm s\in \mathbb{R}^{p+1}$, for some polynomial $P_{a}$ of finite degree $a$ and continuous functions $h_1,\ldots,h_{p+1}$. In the case of mutually independent errors  $\varphi_{ \bm \eta, \eps}(\bm s) =\varphi_{ \eps}(s_{p+1}) + \sum_{k=1}^p\varphi_{\eta_k}(s_k)$ and in the case of jointly normal errors $\varphi_{ \bm \eta, \eps}(\bm s)=P_2(\bm s)$.

Next we define what it means for a random variable to be divisible by a nondegenerate normal distribution and discuss  
how this concept will be used to construct observationally equivalent models.

\medskip
\begin{dfn} \label{dfn:divnorloc}
	A random variable is divisible by a nondegenerate normal distribution if its log characteristic function $\varphi$ can be expressed as $\varphi(s)=\widetilde\varphi(s)-\frac{1}{2}\sigma^2s^2$, for $s \in \mathbb{R}$ and $\sigma^2>0$, where $-\frac{1}{2}\sigma^2s^2$ is the log characteristic function of a nondegenerate zero mean normal distribution and $\widetilde\varphi$ is the log characteristic function of some other distribution.
	Equivalently, 
	the random variable can be expressed as the sum of two independent random variables, one of which is nondegenerate normal.	
\end{dfn}
\medskip

Suppose that $\eps$ is divisible by a nondegenerate normal distribution so that $\eps=\widetilde \eps +Z$ for some independent random variables $\widetilde \eps$ and $Z\sim N(0,\sigma^2)$.
Derivatives of the log characteristic function $\varphi_{\eps}(s)=\varphi_{\widetilde \eps}(s)-\frac{1}{2}\sigma^2s^2$ evaluated at 0 relate the moments of 
$\eps$ and $\widetilde \eps$ producing $E[\eps^j]=E[\widetilde \eps^j]$, for all $j \neq 2$, and $\var(\widetilde\eps)=\var( \eps)-\sigma^2$.
Hence, $\eps$ is divisible by a nondegenerate normal distribution when there exists an $\widetilde \eps$ that has the same moments as $\eps$ except for a smaller variance.

We will construct an observationally equivalent model by assuming that some unobservables are divisible by normal distributions and adding these normal distributions to other unobservables. Equivalently, as suggested in the previous paragraph, we will decrease the variances of some unobservables, increase the variances of other unobservables but leave the observable covariances
unchanged. The assumption of divisibility by a nondegenerate normal distribution guarantees the
validity of decreasing a variance of a random variable leaving it otherwise the same. 

The following assumption is on the validity of transferring nondegenerate normal random variables between the regressors and errors.

\medskip
\begin{asn}\label{as:eiv} 
	There exist $\bm B \in \mathbb{R}^{p \times (p-1)}$ and ${\bm b} \in \mathbb{R}^p$ 
	such that,
	\begin{enumerate}[(i)]
		\item 	 $\bm b'\bm X^* $ is nondegenerate normal,
		\item  $\bm b'\bm X^* $ and $\bm B'\bm X^* $ are independent,
		\item  $(\bm b,\bm B)$ is nonsingular,
		\item if $\cov(\bm \beta' \bm X^*,\bm b'\bm X^*)  < 0$ then $\eps$ is divisible by a nondegenerate normal distribution, 
		\item if $\frac{\cov(X^*_k,\bm b'\bm X^*)}{b_k-\beta_k}< 0$ 
		then $\eta_k$ is divisible by a nondegenerate normal distribution, for $k=1,\ldots,p$.
	\end{enumerate}	
\end{asn}
\medskip

Assumptions \ref{as:eiv}(i)-(iii) state that the regressors can be decomposed into sums of independent components spanning a $p$ dimensional space and one of which is normal.
Assumptions \ref{as:eiv}(iv)-(v) state that if a covariance, depending on $\bm b$, $\bm \beta$, and $\bm X^*$, is negative
then an error is divisible by a normal distribution. 

The assumption is either satisfied for no values of $\bm b$ or infinitely many values of $\bm b$, some of which have arbitrarily small $||\bm b||$, because if $\bm b$ satisfies the assumption then so does $\delta \bm b$, for any $\delta>0$.
Equivalently, after defining $\widetilde {\bm \beta}=\bm \beta - \bm b$,
the assumption is either satisfied for no values of $\widetilde {\bm \beta}$ or infinitely many values of $\widetilde {\bm \beta}$, some of which are arbitrarily close to $\bm \beta$,
because if $\widetilde {\bm \beta}=\bm \beta - \bm b$ satisfies the assumption then so does $\breve{\bm \beta} =\delta \widetilde{\bm \beta} +(1-\delta)\bm \beta=\bm \beta -\delta \bm b$.

Most of the literature identifies the coefficients by assuming that no linear combinations of the regressors are normal or imposes
rank conditions on third or higher moment equations (Assumption \ref{as:eiv}(i) does not hold). 
In the latter case, identification (and estimation) is based on the solution to a system of moment equations.

For any $\bm b$, we provide a $\bm B$
such that $(\bm b,\bm B)$ is nonsingular and $\bm b' \bm X^*$ and $\bm B' \bm X^*$ are uncorrelated.

\medskip
\begin{lma}\label{lma:B}
	Suppose,
	\begin{align*}
	\bm B &= \begin{pmatrix}
	\cov(X^*_2,\bm b'\bm X^*) & \ldots & \cov(X^*_p,\bm b'\bm X^*)\\
	-\cov(X^*_1,\bm b'\bm X^*) & \bm 0 & \bm 0 \\
	\bm 0 & \ddots & \bm 0 \\
	\bm 0 & \bm 0 & -\cov(X^*_1,\bm b'\bm X^*)
	\end{pmatrix}.
	\end{align*}
	If $\var(\bm b'\bm X^*)>0$ and $\cov(X^*_1,\bm b'\bm X^*)\neq 0$ then $(\bm b,\bm B)$ is nonsingular and $\cov(\bm b' \bm X^*,\bm B' \bm X^*)=\bm 0$.	
\end{lma}
\medskip

Even if $\bm b'\bm X^*$ and $\bm B'\bm X^*$ are uncorrelated, they may still be dependent.
\cite{bekker1986comment} provides the example of a two regressor model with $X^*_1 \sim N(0,\sigma_1^2)$ and $X^*_2=(X^*_1)^2-\sigma_1^2$. Then $X^*_1$ is normal (Assumption \ref{as:eiv}(i) holds) but there does not exist a nonsingular matrix $(\bm B,\bm b)$ such that $\bm B'\bm X^*$ and $\bm b'\bm X^*$ are independent (Assumption \ref{as:eiv}(ii) does not hold).
Hence, the model is identified because the regressors cannot be decomposed into the sum of two independent components, one of which is normally distributed. 

Now suppose that $\bm X^*$ is jointly normal and
$(\bm B,\bm b)$ is nonsingular with $\bm B'\bm X^*$ and $\bm b'\bm X^*$ independent (Assumptions \ref{as:eiv}(i)-(iii) hold). 
\cite{hong2003simple} show that the model is still identified if the errors are Laplace because they are not divisible by nondegenerate normal distributions (Assumptions \ref{as:eiv}(iv) and (v) fail).

To motivate the construction of an observationally equivalent model and to show how the covariances in Assumptions \ref{as:eiv}(iv)-(v) are obtained,
the following lemma constructs a model with the same observable covariances as the underlying model in \eqref{eq:main}.

\medskip
\begin{lma}\label{lma:dep} Suppose that \eqref{eq:main} holds and for simplicity assume that $\bm X^*\sim N(\bm 0,\bm \Sigma_{{\bm X}^*})$. 
	Let $\bm b \in \mathbb{R}^p$ with $||\bm b||$ small and define $\widetilde{\bm \beta}=\bm \beta-\bm b$.
	\begin{enumerate}[(i)]
		\item Suppose Assumptions \ref{as:eivdep}(i) and \ref{as:eiv}(iv)-(v) hold.		
		Let $Z_{1k}\sim N(0,|{\cov(X^*_k,\bm b'\bm X^*)}/{\widetilde \beta_k}|)$, $Z'_{1k}\sim N(0,|{\cov(X^*_k,\bm b'\bm X^*)}/{\widetilde \beta_k}|)$, and $Z_2\sim N(0,|\cov(\bm \beta' \bm X^*,\bm b'\bm X^*)|)$.
		If $\cov(X^*_k,\bm b'\bm X^*)/\widetilde \beta_k>0$ then define ${\widetilde X}^*_k= X^*_k +Z_{1k}$ and $ \eta_k= \widetilde \eta_k +Z'_{1k}$ otherwise define $ X^*_k= {\widetilde X}^*_k +Z_{1k}$ and $\widetilde \eta_k =   \eta_k +Z'_{1k}$.	
		If $\cov(\bm \beta' \bm X^*,\bm b'\bm X^*)<0$ then define $ \eps =\widetilde \eps + Z_2$ otherwise define $\widetilde \eps =\eps + Z_2$.
		Then the observable covariances are, 	
		\begin{align*}
		\begin{pmatrix}\bm \Sigma_{\bm X} & \bm \Sigma_{\bm X,Y}'\\ \bm \Sigma_{\bm X,Y} & \sigma^2_Y 
		\end{pmatrix}
		&=\begin{pmatrix}  \bm \Sigma_{\bm X^*} +\bm \Sigma_{\bm \eta} & \bm \beta'\bm \Sigma_{\bm X^*} \\ \bm \Sigma_{\bm X^*}\bm \beta  &\bm \beta' \bm \Sigma_{\bm X^*}\bm \beta +\sigma^2_{\eps}\end{pmatrix}
		=\begin{pmatrix}{\bm \Sigma}_{\widetilde{\bm X}^*} +{\bm \Sigma}_{\widetilde{\bm \eta}} &\widetilde{\bm \beta}'{ \bm \Sigma}_{\widetilde{\bm X}^*}\\ {\bm \Sigma}_{\widetilde{\bm X}^*}\widetilde{\bm \beta}  & \widetilde{\bm \beta}' {\bm \Sigma}_{\widetilde{\bm X}^*} \widetilde{\bm \beta} +\sigma^2_{\widetilde{\eps}}
		\end{pmatrix}.
		\end{align*}
		\item Suppose Assumption  \ref{as:eivdep}(ii) holds.
		Define 
		$(\widetilde{\bm \eta},\widetilde\eps) \sim N(\bm 0,\widetilde{\bm \Sigma})$
		and $\widetilde{\bm X}^* \sim N(\bm 0,{\bm \Sigma}_{ \widetilde{\bm X}^*})$,
		where  $\widetilde{\bm \Sigma} =\bm \Sigma +\bm \Omega$,
		$ {\bm \Sigma}_{\widetilde{\bm X}^*} = \bm \Sigma_{\bm X^*} -\bm \Omega_{\bm \eta}$, 
		and {\small $\bm \Omega	=\begin{pmatrix}  \bm \Omega_{\bm \eta}& \bm \Omega_{\bm \eta,\eps}\\
			\bm \Omega_{\bm \eta,\eps}' &  \Omega_{\eps}
			\end{pmatrix}$}	satisfies	$\bm \Omega_{\bm \eta}\widetilde{\bm \beta} =\bm \Omega_{\bm \eta,\eps} - \bm \Sigma_{\bm X^*} \bm  b $
		and
		$ \Omega_{\eps} 	=\bm \beta '\bm \Sigma_{\bm X^*} \bm b  +  \widetilde{\bm \beta}' \bm \Omega_{\bm \eta,\eps} $. 
		Then the observable covariances are,
		\begin{align*}
		\begin{pmatrix}\bm \Sigma_{\bm X} & \bm \Sigma'_{\bm X,Y}\\ \bm \Sigma_{\bm X,Y} &\sigma^2_Y
		\end{pmatrix}
		&=\begin{pmatrix} \bm \Sigma_{\bm X^*} +\bm \Sigma_{\bm \eta}  & \bm \beta\bm \Sigma_{\bm X^*} +\bm \Sigma_{\bm \eta,\eps}'\\ \bm \Sigma_{\bm X^*}\bm \beta +\bm \Sigma_{\bm \eta,\eps} & \bm \beta' \bm \Sigma_{\bm X^*}\bm \beta +\sigma^2_{\eps}\end{pmatrix}
		=\begin{pmatrix}{\bm \Sigma}_{\widetilde{\bm X}^*} +{\bm \Sigma}_{\widetilde{\bm \eta}}&\widetilde{\bm \beta}'{ \bm \Sigma}_{\widetilde{\bm X}^*} +{\bm \Sigma}_{ \widetilde{\bm \eta},\widetilde\eps}'\\ {\bm \Sigma}_{\widetilde{\bm X}^*}\widetilde{\bm \beta} +{\bm \Sigma}_{\widetilde{\bm \eta}, \widetilde\eps} & \widetilde{\bm \beta}' {\bm \Sigma}_{\widetilde{\bm X}^*} \widetilde{\bm \beta} +\sigma^2_{\widetilde{\eps}} 
		\end{pmatrix}.
		\end{align*}
	\end{enumerate}	
\end{lma}

\medskip

Lemma \ref{lma:dep} constructs infinitely many observationally equivalent models, one for each choice of $\bm b\in \mathbb{R}^p$ as long as $\widetilde{\bm \Sigma}\succ 0$ and $\bm \Sigma_{ \widetilde{\bm X}^*} \succ 0$
\citep[see][]{klepper1984consistent}.
If $||\bm b||$ is small enough (i.e., $\widetilde{\bm \beta}$ is close enough to $\bm \beta$)    
then $\bm \Sigma_{ \widetilde{\bm X}^*}\succ 0$ and $\widetilde{\bm \Sigma}\succ 0$ 
because $\bm \Sigma_{ {\bm X}^*} \succ 0$, ${\bm \Sigma} \succ 0$, and
$||\bm \Omega|| \leq ||\bm b|| \cdot P_0$.
For example, $\widetilde{\bm \Sigma}\succ 0$ implies that $\sigma_{\widetilde \eps}^2=\sigma^2_{\eps}+\bm \beta '\bm \Sigma_{\bm X^*} \bm b  +  \widetilde{\bm \beta}' \bm \Omega_{\bm \eta,\eps}
>0$.
If $\sigma_{\eps}^2$ is small then this can restrict $||\bm b||$ (and $||\bm \Omega_{\bm \eta,\eps}||$) to be small.

The construction in part (i) shifts the variance of $\eps$ by $|\cov(\bm \beta' \bm X^*,\bm b'\bm X^*)|$
and the variance of $\eta_k$ by $|{\cov(X^*_k,\bm b'\bm X^*)}/\widetilde{\beta}_k |$,
and compensates for these shifts with opposing shifts in the variance of $X^*_k$.
Depending on the direction of a shift we either add a nondegenerate normal distribution or assume divisibility by a nondegenerate normal distribution by Assumptions \ref{as:eiv}(iv)-(v).
The construction in part (ii) in the case $ \bm \Omega_{\bm \eta,\eps}=\bm 0$ is the same as part (i) because $\Omega_{\eps}=\cov(\bm \beta' \bm X^*,\bm b'\bm X^*)$,
$\Omega_{\eta_j,\eta_k}=-\cov(X^*_k,\bm b'\bm X^*)/\widetilde \beta_k $ if $k=j$, and $\Omega_{\eta_j,\eta_k}=0$ if $k \neq j$.
The errors are jointly normal by Assumption \ref{as:eivdep}(ii) and hence automatically divisible by normal distributions.

Constructing models that are observationally equivalent to \eqref{eq:main} will be analogous to the constructions in Lemma \ref{lma:dep}. 
The difference is that instead of adding and subtracting variances, we add and subtract quadratic terms to log characteristic functions.

The following theorem establishes identification of the errors-in-variables model with either mutually independent errors
or with jointly normal errors, extending \cite{reiersol1950} from the single regressor model to the multiple regressor model.

 \medskip

\begin{thm}\label{th:eiv} 
	\
	
	\begin{enumerate}[(i)]
		\item Suppose that \eqref{eq:main} and Assumption \ref{as:eivdep}(i) hold. 
		Assume that $E[e^{\lambda|\eps|}]<\infty$ and $E[e^{\lambda|\eta_k|}]<\infty$, $k=1,\ldots,p$, for some $\lambda>0$.		
		The vector $\bm \beta$ is not identified if and only if 
		Assumptions \ref{as:eiv}(i)-(v) hold. 
		\item Suppose that \eqref{eq:main} and Assumption  \ref{as:eivdep}(ii) hold. 
		The vector $\bm \beta$ is not identified if and only if 
		Assumptions \ref{as:eiv}(i)-(iii) hold.	
	\end{enumerate}

\end{thm}
\medskip

We provide a broad outline of the proof leaving the details to Section \ref{ap:eiv} of the appendix. 
The log characteristic function of \eqref{eq:main} is,
\begin{align}
\varphi_{\bm X, Y}(\bm s )
&=\varphi_{\bm X^*}((\bm I_{p}  , \bm \beta)\bm s) 
+\varphi_{ \bm \eta, \eps}(\bm s),  \qquad \bm s \in \mathbb{R}^{p+1},\label{eq:lcf}
\end{align}
where $\varphi_{ \bm \eta, \eps}(\bm s) =\varphi_{ \eps}(s_{p+1}) +\sum_{k=1}^p\varphi_{\eta_k}(s_k)$ is separable when the errors are mutually independent (Assumption \ref{as:eivdep}(i) holds) or $\varphi_{ \bm \eta, \eps}(\bm s)=-\frac{1}{2} \bm s'\bm \Sigma \bm s$ is quadratic when the errors are jointly normal  (Assumption \ref{as:eivdep}(ii) holds).
Sufficient conditions for identification are then obtained by strategically choosing the arguments $\bm s$ (which represent linear combinations of $\bm Y$) and taking derivatives, thus eliminating $\varphi_{ \bm \eta, \eps}(\bm s)$ from the equation.
To prove necessity of these conditions we construct an observationally equivalent model by using Assumptions \ref{as:eiv}(i)-(iii) to decompose $\bm X^*$ into the sum of two independent components, one of which is normal, and  
then transferring the normal component from the regressors to the errors by Assumptions \ref{as:eiv}(iv)-(v) or Assumption \ref{as:eivdep}(ii).

The sufficiency part of the proof analyzes characteristic functions in neighborhoods of the origin (no moments need to exist). The necessity part of the proof, however, constructs an observationally equivalent model by defining log characteristic functions on their domains. 
The necessary and sufficient conditions coincide because the errors (but not necessarily the regressors) have analytic characteristic functions
so that relationships can be extended from neighborhoods of the origin to domains of functions by analytic continuation.\footnote{The assumption that the errors are analytic ($E[e^{\lambda|\eps|}]<\infty$ and $E[e^{\lambda|\eta_k|}]<\infty$) can be replaced by the assumption that the errors are subgaussian or bounded.}$^,$\footnote{\cite{szekely2000identifiability} provide examples where relationships in a neighborhood of the origin do not extend to domains of non-analytic characteristic functions.}

Theorem \ref{th:eiv}(i) in the case of a single regressor does not need the assumption that the errors have analytic characteristic functions \citep{reiersol1950}. 

\medskip
\begin{cor}\label{co:simeiv}
	Suppose that $X=X^*+\eta$ and $Y=\beta X^*+\eps$. 
	Assume that $X^*$, $\eps$, and $\eta$ are mutually independent. 
	The coefficient $\beta$ is not identified if and only if (i) $X^*$ is nondegenerate normal and (ii) either $\eta$ or $\eps$ is divisible by a nondegenerate normal distribution. 
\end{cor}
\medskip
 
The main difference between Theorem \ref{th:eiv}(ii) and \cite{bekker1986comment} is that he assumes $\bm \eta$ and $\eps$ are independent while we
allow $\bm \eta$ and $\eps$ to be arbitrarily dependent. To show necessity, 
\cite{bekker1986comment} sets up a system of equations with  more parameters than equations but does not 
explicitly provide multiple solutions to the system while we explicitly construct an observationally equivalent model.

\section{Estimation}\label{se:eivest}

In this section we mimic the steps from our identification proof to estimate $\bm \beta$ in the errors-in-variables model \eqref{eq:main}. 
The estimator uses a continuum of moment conditions based on second derivatives of the log characteristic function of the observables, 
which has the potential to be more efficient than estimators based on third and higher moments because the characteristic function completely characterizes the distribution of the underlying random variables.
The estimator is root-$n$ consistent and asymptotically normal.

Consider the following second derivatives of the log characteristic  function \eqref{eq:lcf} evaluated at $\bm s= (\bm b,-1)  v$, for $\bm b\in \mathcal{B} \subset \mathbb{R}^{p}$,
 less the derivatives evaluated at $\bm s=\bm 0$,
\begin{align*}
\cov (X_{k_1},Y)
+  \left.\frac{{\partial^{2}	\varphi_{\bm  X,Y}({\bm  s})}}{ \partial s_{k_1} \partial s_{p+1}} \right|_{\bm s= (\bm b,-1)  v}
&=\sum_{k=1}^p \beta_{k}  \left(\cov \left(X^*_{k_1}, X^*_{k}\right)
+  \left. \frac{\partial^2 \varphi_{\bm X^*}(\bm u)}{\partial u_{k_1}\partial u_{k}}   \right|_{\bm u =( \bm b-\bm \beta)	v}
\right), \ 1\leq k_1  \leq p,\\
\cov (X_{k_1},X_{k_2})
+ \left. \frac{{\partial^{2}	\varphi_{\bm  X,Y}({\bm  s})}}{ \partial s_{k_1} \partial s_{k_2}}  \right|_{\bm s= (\bm b,-1)  v}
&=\cov \left(X^*_{k_1}, X^*_{k_2}\right)
+  \left. \frac{\partial^2 \varphi_{\bm X^*}(\bm u)}{\partial u_{k_1}\partial u_{k_2}}   \right|_{\bm u =( \bm b-\bm \beta)	v}
, \quad 1\leq k_1<k_2  \leq p.
\end{align*}
The left sides of the equations are expressed in terms of covariances and second derivatives of the log characteristic function of the observables $(\bm X,Y)$ 
and the arguments $\bm b$.
The right sides of the equations are expressed in terms of covariances and second derivatives of the log characteristic function of the unobservables $\bm X^*$,
the unknown coefficients $\bm \beta$, and the arguments $\bm b$.
We can interpret the above expressions as measures of nongaussianity since they are all identically $0$ for any value of $\bm b$ for which $(\bm b -\bm \beta)'\bm X^*$ is normal. 

The estimator is now based on the following moments,
\begin{align}
\bm H(\bm b,v) &=\left( \cov (X_{1},X_{2})
+ \left. \frac{ {\partial^{2}	\varphi_{\bm  X,Y}({\bm  s})}}{ \partial s_{1} \partial s_{2}} 
\right|_{\bm s =  (\bm b,-1) v},\ldots, \cov (X_{p-1},X_{p})
+ \left. \frac{ {\partial^{2}	\varphi_{\bm  X,Y}({\bm  s})}}{ \partial s_{p-1} \partial s_{p}} 
\right|_{\bm s =  (\bm b,-1) v},
\right. \nonumber \\
&\quad \quad \left.
\cov (X_{1},Y)
+ \left. \frac{{\partial^{2}	\varphi_{\bm  X,Y}({\bm  s})}}{ \partial s_{1} \partial s_{p+1}} 
\right|_{\bm s =  (\bm b,-1) v},\ldots,
\cov (X_{p},Y)
+ \left. \frac{ {\partial^{2}	\varphi_{\bm  X,Y}({\bm  s})}}{ \partial s_{p} \partial s_{p+1}} 
\right|_{\bm s =  (\bm b,-1) v} \right)', \label{eq:H}
\end{align}
where $\bm H(\bm b,v)'$ is the conjugate transpose of $\bm H(\bm b,v)\in \mathbb{C}^{p(p+1)/2\times 1}$.

We impose the identifying assumption that no linear combinations of $\bm X^*$ are normal (i.e., Assumption \ref{as:eiv}(i) does not hold). 

\medskip
\begin{asn} \label{as:nongaus}
	There does not exist a nonzero $\bm b \in \mathcal{B}$ such that $\bm b' \bm X^*$ is normal. 
\end{asn}
\medskip

If Assumption \ref{as:nongaus} holds then $\bm \beta$ is identified from the moments $\bm H(\bm b,v)=\bm 0$
(i.e., $\bm H(\bm b,v) =\bm 0$ if and only if $\bm b=\bm \beta$).
Variation in $\frac{ {\partial^{2}	\varphi_{\bm  X,Y}({\bm  s})}}{ \partial s_{k_1} \partial s_{k_2}} $ implies that $(\bm X,Y)$ is not jointly normal and is a necessary condition for the assumption to hold.

Now to estimate the coefficients using the moments $\bm H(\bm b,v)$,
we can fix $v$ and choose how to weight each of the moments based on minimizing a norm.
Then, we can choose how to weight $v\in \mathbb{R}$ based on minimizing another norm.
\cite{carrasco2000generalization} and \cite{carrasco2017efficient} analyze the latter of these choices taking the former as given.
They show that the optimal weight on a continuum of moment conditions is the inverse of an asymptotic covariance operator in Hilbert space. 
This inverse is discontinuous and so its estimation requires some regularization.
In this setting, only a few papers analyze which regularization technique to use and how to choose regularization parameters to estimate the optimal weights.
Further, estimates are usually sensitive to the regularization technique and the choice of regularization parameters (which can be data intensive). 
We opt for the simpler IGMM estimator, which uses the optimal GMM weight matrix for each $v$ and exponentially decaying weight on $v$,
\begin{align}
\widehat {\bm \beta} &=  
\argmin_{\bm b \in \mathcal{B}}  \int_{-\infty}^{\infty}\bm H_n(\bm b,v)'\bm W_n(v)\bm H_n(\bm b,v) \pi(v) dv,  \label{eq:IGMM}
\end{align}
where $\pi(v)={e^{-v^2/2}}/{\sqrt{2\pi}}$ is the density of a standard normal distribution, 
$\bm H_n(\bm b,v)$ is the natural sample analog of $\bm H(\bm b,v)$,
$\bm W_n(v)$ is a random positive definite matrix approaching the optimal GMM matrix weight $\bm \Omega(v,v)^{-1}$,
where $\sqrt{n}\bm H_n(\bm \beta,v) \stackrel{p}{\rightarrow} N(\bm 0, \bm \Omega(v,v))$, and
$\bm H_n(\bm b,v)'\bm W_n(v)\bm H_n(\bm b,v)$ is the optimal GMM norm (see Section \ref{ap:asym} of the appendix for further details).
Exponentially decaying weights 
reflect that the characteristic function is more accurately estimated closer to the origin and contains more information about the underlying distribution closer to the origin. They are commonly used in estimation of parameters using characteristic functions \citep[see][]{yu2004empirical}.

Next we assume random sampling and impose regularity conditions.

\medskip
\begin{asn} \label{as:iid}
	We observe a sample $\{\bm X_{i},Y_i\}_{i=1}^n$ of independently and identically distributed 
	draws from the distribution of $(\bm X,Y)$.
\end{asn}
\medskip

\begin{asn} \label{as:reg} \
	\begin{enumerate}[(i)]
		\item  $\bm \beta$ belongs to the interior of the compact set $\mathcal{B}$,
		\item 	$E[e^{\lambda |Y|}]< \infty$ and $E[e^{\lambda|X_k|}]< \infty$, $k=1,\ldots,p$, for some $\lambda>0$,
		\item  $E[e^{\textup{i}v({\bm b}' \bm X-Y) }] \neq 0$ for all $\bm b\in\mathcal{B}$ and $v$ almost everywhere, 
		\item  $ \bm W_n(v) \stackrel{p}{\rightarrow} \bm W(v)$ almost everywhere for $\bm W(v)$ positive definite,
		\item  $\bm H(\bm b,v)'\bm W(v)\bm H(\bm b,v) \in {L}^2(\pi)$ and all elements of
		$\frac{\partial \bm H(\bm b,v)}{\partial \bm b}\bm W(v)\bm H(\bm b,v) \in {L}^2(\pi)$, for all $\bm b \in \mathcal{B}$,
		where ${L}^2(\pi)=\left\{g:\mathbb{R} \rightarrow \mathbb{R}\ | \ \int g(v)g(v)' \pi(v)dv <\infty \right\}$, 
		\item   $\int_{-\infty}^{\infty}\frac{\partial \bm H}{\partial \bm b}(\bm \beta,v)\bm W(v)\frac{\partial \bm H}{\partial \bm b'}( \bm \beta,v) \pi(v) dv$ and
		$\int_{-\infty}^{\infty} \int_{-\infty}^{\infty}
		\frac{\partial \bm H}{\partial \bm b}(  {\bm \beta},v)\bm W(v)\bm \Omega(v,w)\bm W(w) \frac{\partial \bm H}{\partial \bm b'}(  {\bm \beta},w)\pi(v)\pi(w)dvdw$ are nonsingular and finite.
	\end{enumerate}
	
\end{asn} 
\medskip

Using Assumption \ref{as:nongaus} to show identification, 
a random sample by Assumption \ref{as:iid},  
and Assumption \ref{as:reg}'s regularity conditions, 
the following theorem 
establishes that $\widehat {\bm \beta}$ is root-$n$ asymptotically normal.

\medskip

\begin{thm} \label{th:asym}
	Suppose that \eqref{eq:main} holds, either Assumption \ref{as:eivdep}(i) holds or Assumption \ref{as:eivdep}(ii) holds, and Assumptions \ref{as:nongaus},  \ref{as:iid}, and \ref{as:reg} with $\bm W(v)=(\bm \Omega(v,v))^{-1}$ hold. Then,
{\small\begin{align*}
\sqrt{n}(\widehat {\bm \beta}-\bm \beta)
&\stackrel{d}{\rightarrow}N\left(\bm 0,\bm S \left(\int_{-\infty}^{\infty} \int_{-\infty}^{\infty}
\frac{\partial \bm H}{\partial \bm b}(  {\bm \beta},v)\bm \Omega^{-1}(v,v)\bm \Omega(v,w)\bm \Omega^{-1}(w,w) \frac{\partial \bm H}{\partial \bm b'}(  {\bm \beta},w)\pi(v)\pi(w)dvdw
\right)
\bm S'\right), 
\end{align*}
}where 
$
\bm S = \left( \int_{-\infty}^{\infty}\frac{\partial \bm H}{\partial \bm b}( {\bm \beta},v)\bm \Omega^{-1}(v,v)\frac{\partial \bm H}{\partial \bm b'}(  {\bm \beta},v) \pi(v) dv\right)^{-1}
$
and $\bm \Omega(v,w)$ is defined in Section \ref{ap:asym} of the appendix. 
\end{thm}
\medskip

\section{Monte Carlo simulations}\label{se:eivsim}

In this section we 
compare the finite sample performance of the IGMM estimator to the OLS estimator and a GMM estimator based on third to fifth moments allowing for various amounts of measurement error and for some mismeasured and perfectly measured regressors. The IGMM and GMM estimators produce results that have small bias and decreasing standard deviation at root-$n$ rate. These estimators are robust to measurement error and the number of mismeasured regressors. 
When there is measurement error the OLS estimator produces results that are severely biased with zero coverage probability of the 95\% confidence interval. 

The data $\{\bm X_{i},\bm Z_{i},Y_i\}_{i=1}^n$,  for $n \in \{500, 1000, 2000\}$, is simulated $100$ times from the errors-in-variables model, 
\begin{align}
\left\{
\begin{array}{rl}
\bm X &=\bm X^* +\bm \eta,\\
Y &= \bm \beta'\bm X^* +\bm \gamma'\bm Z +\eps,
\end{array}
\right.
\label{eq:sim} 
\end{align}
where $(\bm X^*,\bm Z)$, $\bm \eta$, and $\eps$ are mutually independent.
The vector $\bm X^*$ contains unobserved mismeasured regressors and the vector $\bm Z$ contains perfectly measured observed regressors.
The regressors $(\bm X^*,\bm Z)$ are multidimensional $\chi^2$ distributed with $10$ degrees of freedom and covariance matrix adjusted to have diagonal elements (variances) equal to 3 and off-diagonal elements (covariances) equal to 1, 
the structural error $\eps$ is standard normal,
the measurement errors $\bm \eta$ are either identically zero (no measurement errors),
independent normal with variance $1$ (independent measurement errors),
or jointly normal with covariance matrix having diagonal elements equal to 1 and off-diagonal elements equal to 1/2 (dependent measurement errors).

The parameters in $\bm \beta $ and $\bm \gamma$ are all equal to 1.
Denote the dimension of $\bm X$ (and $\bm \beta$) by $N_x$ and the dimension of $\bm Z$ (and $\bm \gamma$) by  $N_z$, and
let $N_x=1,2,3$ and $N_x+N_z=4$. 
The estimators all use the preliminary step of partialling out the perfectly measured regressors $\bm Z$.
We compare three estimators that use the residuals of the regressions of $Y$ on $\bm Z$ and $\bm X$ on $\bm Z$: 
\begin{enumerate}[(a)]

	\item the IGMM estimator in \eqref{eq:IGMM} with starting values for the optimization coming from the GMM estimates,\footnote{The estimator produces similar results when the starting values for the optimization are the GMM estimates, the OLS estimates or the underlying coefficient values.}
		\item the GMM estimator based on third, fourth and fifth  moments \citep[see][]{erickson2017xtewreg},\footnote{The estimates use the STATA function ``xtewreg'' available on the authors' websites.} and
 \item the OLS estimator that ignores measurement error.
\end{enumerate}

\medskip

The integral for the IGMM estimator is estimated by $\sum_{j=1}^{50} \bm H_n(\bm b,v_j)'\bm W_n(v_j)\bm H_n(\bm b,v_j)$, where $v_1,\ldots,v_{50}$ are $50$ independent draws from a standard normal distribution,
$\bm H_n(\bm b,v_j)$ is the sample analog of $\bm H(\bm b,v_j)$ in \eqref{eq:H}, and $\bm W_n(v_j)$ is the sample analog of $(\bm \Omega(v_j,v_j))^{-1}$ in \eqref{eq:Omegaterms} with the unknown $\bm \beta$ replaced by the GMM estimator.\footnote{The results are similar when using the identity matrix as the weight matrix.} 
The weights $\pi(v)={e^{-v^2/2}}/{\sqrt{2\pi}}$ decay rapidly and are not data-driven.
The estimates are sensitive to these weights and we view the choice of the weights as similar to the choice of bandwidth in other nonparametric settings like deconvolution. 
By allowing the weights to depend on the data, the estimator could take into account how information decays as $|v|$ grows and how well the characteristic function is estimated as $|v|$ grows. Incorporating these weights into the estimator as in \cite{carrasco2000generalization} has the potential to improve the IGMM estimator even further (see also the discussion surrounding \eqref{eq:IGMM}).

We plot the sample analogs of $ \frac{{\partial^2	\varphi_{ {\bm X}, Y}(\bm  s)}}{ \partial s_{k_1}\partial s_{k_2}} $
to confirm variation in these second derivatives as suggestive evidence that no linear combinations of the unobserved regressors are normal.

Using $100$ simulations each of sample size $n \in \{500,1000,2000\}$, Tables 1 and 2 report the bias, standard deviation (std), and coverage probability (cp) of the 95\% confidence interval of the estimates for $\beta_1=1$  in Model \eqref{eq:sim}. The confidence intervals use $50$ bootstrapped samples to estimate standard errors. Table 1 displays the results when $N_x=N_z=2$ and for various amounts of measurement errors and Table 2 displays the results for a varying number of mismeasured and perfectly measured regressors.  


When there is no measurement error, the OLS estimator is unbiased, has smallest standard deviations, and has coverage probabilities that are close to the nominal level of 95\%.
However, when there is measurement error then the OLS estimator has large bias that does not improve with sample size. 
The standard deviations are deceptively small and the coverage probabilities are zero.
The OLS estimator has incorrect rejection rates and coverage probabilities and inference based on this estimator will not be trustworthy.

The IGMM and GMM estimators perform well in all the simulations. 
They have small bias and standard deviations that decrease at the root-$n$ convergence rate in Theorem \ref{th:asym}. The coverage probability is close to the nominal level for the IGMM estimator but sometimes far away for the GMM estimator, suggesting that inference using the GMM estimator might not be trustworthy.
Table 1 shows that the bias and standard deviation increase with measurement error but remain relatively stable.
Table 2 shows that the number of mismeasured regressor increases bias by a little but has a negligible effect on the standard deviation.
We experimented with different distributional choices and covariance matrices for the regressors and errors, producing results that were qualitatively similar to the ones presented in Tables 1 and 2.

{\linespread{1.30}
\captionsetup{width=.85\textwidth}

\begin{table}[H]
	{\small
	\caption{\small Performances of the IGMM, GMM, and OLS estimators 
		with $N_x=N_z=2$ and 
		various amounts of measurement errors}
	\vspace{-0.5em}
		\begin{center}
			\begin{tabular}{llccccccccc}
				\hline\hline
				& 	&   \multicolumn{3}{c}{$(\sigma_{\eta}^2,\sigma_{\eta_1,\eta_2})=(0,0)$}  & \multicolumn{3}{c}{$(\sigma_{\eta}^2,\sigma_{\eta_1,\eta_2})=(1,0)$}& \multicolumn{3}{c}{$(\sigma_{\eta}^2,\sigma_{\eta_1,\eta_2})=(1,1/2)$}  \T \\
				n&& IGMM& GMM & OLS& IGMM& GMM & OLS& IGMM& GMM & OLS   \T \\ \hline\hline
				500 
				&bias & 0.02 & 0.01 & 0.00 & -0.06 & -0.10 &-0.25 &-0.11 &-0.15 &-0.33 \\ 
				&std   & 0.10 & 0.09&0.03 &  0.15  &0.16 &0.03 &0.16 &0.19 &0.04 \\
				&cp   &0.90 &0.72 &0.92   &0.94 &0.64 &0.00 &0.92 &0.59 &0.00 \\   \hline
				1000
				&bias  &0.01 & -0.01&0.00 &-0.05 &-0.05 & -0.25&-0.08 &-0.08 &-0.33 \\ 
				&std  &0.07 & 0.06&  0.02& 0.13& 0.13&0.03 &0.13 &0.14 &0.03 \\ 
				&cp   &0.95 &0.83 &0.97 &0.88 &0.73 &0.00 &0.89 &0.68 &0.00 \\   \hline
				2000
				&bias  &0.00 & 0.00&0.00  &-0.04& -0.03& -0.25 &-0.06 &-0.05 & -0.33\\  
				&std  &0.05 & 0.04&0.02   &0.09 &0.09 &0.02 &0.09 &0.09 &0.02\\ 
				&cp    &0.95 &0.84 &0.93  &0.91 &0.77 &0.00 &0.90 &0.79 &0.00 \\   \hline\hline
				\multicolumn{11}{p{420pt}}{\footnotesize Notes: Bias, standard deviation, and coverage probability (cp) of the 95\% confidence interval are estimated from $100$ simulations each with sample size $n$. 
				The confidence intervals use 50 bootstrapped samples to estimate standard errors.
				The measurement errors are jointly normal with covariance matrix having diagonal elements 	$\sigma_{\eta}^2$ and off-diagonal elements $\sigma_{\eta_1,\eta_2}$.
			}
			\end{tabular}
		\end{center}
	}
\end{table}

\vspace{-0.5em}

\captionsetup{width=.85\textwidth}

\begin{table}[H]
	{\small
	\caption{\small Performances of the IGMM, GMM, and OLS estimators 
	with a varying number of mismeasured regressors $N_x$ and perfectly measured regressors $N_z$
	}
	\vspace{-0.5em}
		\begin{center}
			\begin{tabular}{llccccccccc}
				\hline\hline
				 & &   \multicolumn{3}{c}{$(N_x,N_z)=(1,3)$}  & \multicolumn{3}{c}{$(N_x,N_z)=(2,2)$}& \multicolumn{3}{c}{$(N_x,N_z)=(3,1)$}  \T \\
				n&& IGMM& GMM & OLS& IGMM& GMM & OLS& IGMM& GMM & OLS   \T \\ \hline\hline
				500
					&bias  &-0.03 &-0.06 & -0.29&-0.06 &-0.10 &-0.25 &-0.08 &-0.11 &-0.20 \\  
					&std   &0.18 &0.15 &0.03    &0.15 &0.16 &0.03 &0.14 &0.16 &0.04 \\  
					&cp    &0.95 &0.77 &0.00    &0.94 &0.64 &0.00 &0.90 &0.43 &0.00 \\   \hline
				1000
				&bias  &-0.03 & -0.02&-0.29 &-0.05 &-0.05 & -0.25&-0.07 &-0.07 &-0.20 \\ 
				&std   &0.15 & 0.11&0.02    & 0.13& 0.13&0.03 &0.13 &0.12 &0.03 \\ 
				&cp    &0.90 &0.81 &0.00    &0.88 &0.73 &0.00 &0.88 &0.59 &0.00 \\   \hline
				2000 
				&bias  &-0.02 &0.01 &-0.29 &-0.04 &-0.03 &-0.25 &-0.05 &-0.06 &-0.20 \\   
				&std   &0.11 &0.06 &0.01   &0.09 &0.09 &0.02    &0.09 &0.08 &0.20 \\  
				&cp    &0.92 &0.90 &0.00   &0.91 &0.77 &0.00    &0.89 &0.74 &0.00 \\   \hline \hline
				\multicolumn{11}{p{420pt}}{\footnotesize Notes: Bias, standard deviation, and coverage probability (cp) of the 95\% confidence interval are estimated from $100$ simulations each with sample size $n$. 
					The confidence intervals use 50 bootstrapped samples to estimate standard errors.
					The 
					measurement errors are independent standard normals.
				}
			\end{tabular}
		\end{center}
	}
\end{table}
\vspace{-0.5em}

}

\section{Application}\label{se:app}

Investment is a large component of a country's GDP (for example, investment accounts for about 15\% of U.S. GDP) and hence is central to our understanding
of economic activity \citep[see, e.g.,][]{ABEL1990725}.
Despite an elegant theory for firm investment decisions and extensive empirical research, 
some key variables and/or mechanisms that drive these investment decisions have not been identified. 

The neoclassical theory of firm investment under convex adjustment costs implies that firm investment should be a function only of marginal returns to capital.  
The intuition is that a firm should increase capital when its value is greater than its cost and decrease capital
when its value is less than its cost \citep{keynes1936general}.

Consider the linear regression model,
\begin{equation}
Y = \beta  q^*   + \eps,  \label{eq:tobinQ1}
\end{equation} 
where $Y$ is firm investment, $q^*$ is marginal returns to capital, and $\eps$ is an error.

Empirical studies that regress investment on observed measures of marginal returns to capital explain little about investment decisions.
This rejection of the theory has been interpreted as a sign that financial constraints should be included in the model,
\vspace{-0.5em}
\begin{align}
Y =   \beta_1 q^*  +  \beta_2 FC^*   + \eps,  \label{eq:tobinQ2}
\end{align} 
where $FC^*$ is financial constraints.

Another common explanation for the theory's failure is that marginal returns to capital are measured with error, 
\vspace{-0.5em}
\begin{align}
q =  q^*  + \eta_q,  \label{eq:q}
\end{align}
where $q$ is the ratio of market to book value of equity plus liabilities, and $ \eta_q$ is measurement error.

We investigate the possibility that financial constraints 
are measured with error,
\vspace{-0.5em}
\begin{align}
FC =  FC^*  + \eta_{FC}, \label{eq:FC}
\end{align} 
where $FC$ is cash flows, often used as a measure for financial constraints \citep[see, e.g.,][]{gilchrist1995evidence,cao2019financial}, and $ \eta_{FC}$ is measurement error. 

We use data from \citet{erickson2014minimum} for the years 1992-1995.\footnote{\citet{erickson2014minimum} collect the data 
	from the 2012 Compustat Industrial Files.
	There are $2 850$, $3 031$, $3 279$, and $3 425$ observations in $1992$, $1993$, $1994$, and $1995$, respectively.} 
There is a lot of variation
in the second derivatives of the empirical log characteristic function.
This is strong evidence that the observables are nongaussian and is a necessary condition for no linear combinations of the unobserved regressors to be normal (and hence identification of the coefficients in the models).\footnote{\cite{erickson2014minimum} provide evidence for nongaussianity by showing that the observed data is highly skewed.}

Table 3 compares the IGMM, GMM, and OLS estimators in three models,
with standard errors of the IGMM estimator calculated from 100 bootstrapped samples. 
Model 1 is based on \eqref{eq:tobinQ1} and \eqref{eq:q} 
and excludes a variable for financial constraints, 
Model 2 is based on \eqref{eq:tobinQ2} and \eqref{eq:q}
and assumes that financial constraints are perfectly measured,
and Model 3 is based on  \eqref{eq:tobinQ2}, \eqref{eq:q}, and \eqref{eq:FC}
and allows both marginal returns to capital and financial constraints to have measurement errors.


Ignoring measurement errors, the OLS estimates of the effects of marginal returns to capital and financial constraints on investment decisions are always positive and significant at the 5\% level. 
This supports the majority of the literature that both marginal returns to capital and financial constraints are important determinants of investment decisions.

In Model 1, which excludes financial constraints, the IGMM and GMM estimates are similar, and agree that
the effects of marginal returns to capital on investment decisions are positive and significant.
These estimates are larger than the OLS estimates, which suffers from attenuation bias.

In Model 2, which assumes that financial constraints are perfectly measured,
the IGMM and GMM estimates of the coefficients on marginal returns to capital are all positive, significant,
and similar. The estimates of the coefficients on financial constraints are all insignificant. 
Hence, allowing for measurement error in marginal returns to capital restores the validity of the neoclassical theory that returns to capital, and not financial constraints, are the main drivers of investment decisions \citep[see][]{erickson2000measurement}.

In Model 3, which allows for measurement errors in both marginal returns to capital and financial constraints,
the IGMM and GMM estimates of the coefficients are mostly insignificant. 
This suggests that the theory is missing an important component, or that the linear regression model allowing for measurement errors in both the regressors
is inappropriate due to nonlinearities, or that the regressors and unobservables are statistically dependent.
One possible direction to pursue is a nonlinear model that 
allows for measurement errors in both marginal returns to capital and financial constraints \citep[see, e.g.,][]
{SchennachHu13}.

{\linespread{1.30}
\captionsetup{width=0.60\textwidth}
\begin{table}[H]
	\caption{
			\small 
		Estimates of firm investment decisions based on marginal returns to capital and financial constraints.}
	{\small
		\begin{center}
			\begin{tabular}{lcccccc}
				\hline\hline
				&     \multicolumn{3}{c}{$\beta_1$ (returns to capital)} & \multicolumn{3}{c}{$\beta_2$ (financial constraints)} \\
				Year   & OLS & GMM & IGMM & OLS & GMM & IGMM \rule{0pt}{2.5ex} \\ \hline 
				
				\multicolumn{7}{c}{Model 1: $Y= \beta_1 q^*  + \eps$,
					$q=   q^*  + \eta_q $} \rule[-1.4ex]{0pt}{0pt} \rule{0pt}{2.7ex}  \\		\hline 
				1992 &  0.014$^{**}$ & 0.015$^{**}$ & 0.029$^{**}$ & \multicolumn{3}{c}{\multirow{2}{*}{---}} \\
				& (0.001)  & (0.002) & (0.008) &  \\ 
				1993
				&  0.016$^{**}$ & 0.029$^{**}$ & 0.030$^{**}$ & \multicolumn{3}{c}{\multirow{2}{*}{---}} \\
				& (0.001)  & (0.008) & (0.006) &  \\ 
				1994
				&  0.017$^{**}$ & 0.036$^{**}$ & 0.032$^{**}$ & \multicolumn{3}{c}{\multirow{2}{*}{---}} \\
				& (0.002) & (0.012) & (0.009) &  \\ 
				1995
				&  0.017$^{**}$ & 0.075$^{*}$ & 0.083$^{**}$ & \multicolumn{3}{c}{\multirow{2}{*}{---}} \\
				& (0.002)  & (0.037) & (0.023) &  \\ 	\hline
				
				\multicolumn{7}{c}{Model 2: $Y=  \beta_1 q^*  +\beta_2FC^* + \eps$,
					$q=   q^*  + \eta_q $} \rule[-1.4ex]{0pt}{0pt} \rule{0pt}{2.7ex}  \\
				\hline						
				1992   &  0.012$^{**}$ & 0.017$^{**}$ & 0.029$^{**}$ & 0.077$^{**}$ & 0.056 & -0.001\\
				& (0.002) & (0.007)  & (0.005)  & (0.019) & (0.043) & (0.032) \\ 
				1993 &  0.013$^{**}$ & 0.039$^{**}$ & 0.034$^{**}$ & 0.064$^{**}$ & -0.047 & -0.026 \\
				& (0.002) & (0.008)    & (0.008)     & (0.012)       & (0.037) & (0.031)\\ 
				
				1994 &  0.012$^{**}$ & 0.040$^{**}$ & 0.051$^{**}$ & 0.111$^{**}$          & -0.017 & -0.070 \\
				& (0.002) & (0.017)       & (0.015)  & (0.016)              & (0.078) & (0.077) \\ 
				
				1995 &  0.014$^{**}$ & 0.055$^{**}$ & 0.120$^{**}$ & 0.106$^{**}$ & -0.023  & -0.215 \\
				& (0.002)  & (0.025)  & (0.032)  				& (0.016)        & (0.082) & (0.118) \\ 	\hline
				
				\multicolumn{7}{c}{Model 3: $\begin{array}{l} Y=  \beta_1 q^*  +\beta_2FC^* + \eps, \\ q=   q^*  + \eta_q, \
					FC = FC^*  + \eta_{FC} \end{array}$} \rule[-1.4ex]{0pt}{0pt} \rule{0pt}{2.7ex}  \\
				\hline
				
				1992
				&  0.012$^{**}$ & 0.011$^{**}$ & 0.017$^{**}$ & 0.077$^{**}$ & -0.119$^{**}$ & 0.082 \\
				& (0.002) & (0.003)  & (0.006)  & (0.019) & (0.060) & (0.147) \\ 
				
				1993
				&  0.013$^{**}$ & 0.008 & 0.027$^{**}$ & 0.064$^{**}$ & 0.029 & -0.003 \\
				& (0.002) & (0.009)  & (0.010)  & (0.012) & (0.141) & (0.167) \\ 
				
				1994
				&  0.012$^{**}$ & 0.003 & -0.008 & 0.111$^{**}$ & 0.138   & 0.629$^{**}$\\
				& (0.002) & (0.007)  & (0.012)  & (0.016)   & (0.142) & (0.200) \\ 
				
				1995
				&  0.014$^{**}$ & 0.004 & 0.010 & 0.106$^{**}$ & 0.061   & 0.386 \\
				& (0.002)  & (0.008)  & (0.025)  & (0.016)    & (0.129) & (0.236) \\ 
				\hline \hline
				\multicolumn{7}{p{305pt}}{\footnotesize Notes: Standard errors are 
					in parentheses under the estimate. The IGMM standard errors use 100 bootstrapped samples.
					Double asterisks indicates significance at the 5\% level.}
			\end{tabular}
		\end{center}
	}
\end{table}

\vspace{-0.5em}

}

\section{Conclusion}

We provide necessary and sufficient conditions for identification of the coefficients in the linear errors-in-variables model.
This extends \cite{reiersol1950} from a linear errors-in-variables model with a single mismeasured regressor to multiple mismeasured regressors.
We mimic the identification procedure to produce a new estimator for the coefficients using a continuum of moments based on second derivatives of the log characteristic function of observables.
Monte Carlo simulations 
provide evidence that the estimator has potential to perform well in practice and could be a competitive alternative to already available estimators.  
We apply the estimator to data on firm investment decisions 
and show that the relationship between investment and the regressors, marginal returns to capital and financial constraints, depends on whether or not the regressors contain measurement errors.

\pagebreak

\bibliography{EIVarxiv}

\pagebreak

\appendix

\section{Proofs}

\subsection{Proof of Lemma \protect\ref{lma:B}} \label{ap:B}
The matrix $(\bm b, \bm B)$ is nonsingular because,
\begin{align*}
\det(\bm b,\bm B) 
&=(-1)^{p-1}(\cov(X^*_1,\bm b'\bm X^*))^{p-2}\var(\bm b'\bm X^*)\neq 0,
\end{align*}
where the inequality follows by the assumptions that $\var(\bm b'\bm X^*)>0$ 
and $\cov(X^*_1,\bm b'\bm X^*)\neq 0$. Next,
\begin{align*}
\cov(\bm b' \bm X^*,\bm B' \bm X^*)&=\begin{pmatrix}
\cov(X^*_2,\bm b'\bm X^*)\cov( X^*_1,\bm b' \bm X^*)-\cov( X^*_1,\bm b' \bm X^*)\cov(X^*_2,\bm b'\bm X^*)\\
\vdots \\
\cov(X^*_p,\bm b'\bm X^*)\cov( X^*_1,\bm b' \bm X^*)-\cov( X^*_1,\bm b' \bm X^*)\cov(X^*_p,\bm b'\bm X^*)
\end{pmatrix}=\bm 0,
\end{align*}	
where the first equality follows by substituting in for $\bm B$.
	
\subsection{Proof of Lemma \protect\ref{lma:dep}} \label{ap:lma}
\begin{enumerate}[(i)]
	\item Denote $d_k=\cov(X^*_k,\bm b'\bm X^*)=\sum_{j=1}^pb_j\sigma_{X^*_kX^*_j} $. 
	Then $\cov(\bm \beta'\bm X^*,\bm b'\bm X^*)=\sum_{k=1}^p d_k\beta_k $, 
	$ \sigma_{\widetilde \eps}^2 =\sigma_{\eps}^2+\sum_{k=1}^p d_k \beta_k$, $\sigma_{\widetilde \eta_k}^2=  \sigma_{ \eta_k}^2 -\frac{d_k}{\widetilde \beta_k}$, 
$\sigma_{{\widetilde X}^*_k}^2= \sigma_{X^*_k}^2 +\frac{d_k}{\widetilde \beta_k} $, and $ \sigma_{{\widetilde X}^*_k{\widetilde X}^*_j}=\sigma_{X^*_kX^*_j}$.
Then the observable variances and covariances are,
\begin{align*}
\sigma_{X_k}^2 &= \sigma_{{\widetilde X}^*_k}^2+\sigma_{\widetilde \eta_k}^2
= (\sigma_{X^*_k}^2 +\frac{d_k}{\widetilde \beta_k})+(\sigma_{ \eta_k}^2 -\frac{d_k}{\widetilde \beta_k})=\sigma_{X^*_k}^2+\sigma_{ \eta_k}^2,\\
\sigma_{X_k,X_j}&= \sigma_{{\widetilde X}^*_k{\widetilde X}^*_j}= \sigma_{X^*_kX^*_j},\\
\sigma_{X_j,Y}&=\sum_{k=1}^p\widetilde\beta_k  \sigma_{{\widetilde X}^*_k{\widetilde X}^*_j}
=\sum_{k \neq j}\widetilde\beta_k  \sigma_{X^*_kX^*_j}
+ (\widetilde\beta_j \sigma_j^2 +d_k)
=\sum_{k =1}^p\widetilde\beta_k  \sigma_{X^*_kX^*_j}
+\sum_{j=1}^p b_j\sigma_{X^*_kX^*_j}
=\sum_{k =1}^p\beta_k  \sigma_{X^*_kX^*_j},\\
\sigma_Y^2&=\sum_{k=1}^p\sum_{j=1}^p \widetilde\beta_k\widetilde\beta_j\sigma_{{\widetilde X}^*_k{\widetilde X}^*_j}+\widetilde \sigma_{\eps}^2
=\sum_{k=1}^p \widetilde\beta_k^2(\sigma_{X^*_k}^2 +\frac{d_k}{\widetilde \beta_k})
+\sum_{k\neq j} \widetilde\beta_k\widetilde\beta_j\sigma_{X^*_kX^*_j}
+(\sigma_{\eps}^2+\sum_{k=1}^p d_k\beta_k)\\
&=\sum_{k=1}^p \widetilde\beta_k^2\sigma_{X^*_k}^2 +\sum_{k=1}^p \sum_{j=1}^p\widetilde\beta_k b_j\sigma_{X^*_kX^*_j}
+\sum_{k\neq j} \widetilde\beta_k\widetilde\beta_j\sigma_{X^*_kX^*_j}
+(\sigma_{\eps}^2+\sum_{k=1}^p \sum_{j=1}^p b_j \beta_k\sigma_{X^*_kX^*_j})\\
&=\sum_{k=1}^p (\widetilde\beta_k^2+b_k\beta_k+\widetilde\beta_k b_k)\sigma_{X^*_k}^2 
+\sum_{k\neq j}(\widetilde\beta_k b_j+ \widetilde\beta_k\widetilde\beta_j+b_k \beta_j)\sigma_{X^*_kX^*_j}
+ \sigma_{\eps}^2\\
&=\sum_{k=1}^p\sum_{j=1}^p \beta_k\beta_j \sigma_{X^*_kX^*_j}+ \sigma_{\eps}^2.
\end{align*}
Also, $ \sigma_{\widetilde \eps}^2>0$,  $\sigma_{\widetilde \eta_k}^2>0$, and $\bm \Sigma_{\widetilde{\bm X}^*}\succ 0$ for small enough $||\bm b||$
because $ \sigma_{\eps}^2>0$, $\sigma_{ \eta_k}^2>0$, $\bm \Sigma_{{\bm X}^*} \succ 0$, and $|d_k|\leq ||\bm b||\cdot ||\bm \Sigma_{{\bm X}^*}  ||$.

\item Using $\bm  \Omega_{\bm \eta}\widetilde{\bm \beta} = \bm \Omega_{\bm \eta,\eps}-\bm \Sigma_{\bm X^*}  \bm b$ 
and $ \Omega_{\eps} =\bm \beta'\bm \Sigma_{\bm X^*}  \bm b+ \widetilde{\bm \beta}' \bm \Omega_{\bm \eta,\eps} $ we obtain the observable variances and covariances, 
\begin{align*}
\bm \Sigma_{\bm X} &=  {\bm \Sigma}_{\widetilde{\bm X}^*} +  {\bm \Sigma}_{\widetilde{\bm \eta}}
= (\bm \Sigma_{\bm X^*} -\bm \Omega_{\bm \eta})+ (\bm \Sigma_{\bm \eta} +\bm \Omega_{\bm \eta})=\bm \Sigma_{\bm X^*}+\bm \Sigma_{\bm \eta},\\
\bm \Sigma_{\bm X,Y}&= {\bm \Sigma}_{\widetilde{\bm X}^*} \widetilde{\bm \beta} +  {\bm \Sigma}_{\widetilde{\bm \eta},\widetilde\eps}
=(\bm \Sigma_{\bm X^*} -\bm \Omega_{\bm \eta})\widetilde{\bm \beta}+ (\bm  \Sigma_{\bm \eta,\eps}+\bm \Omega_{\bm \eta,\eps})\\
&=\bm \Sigma_{\bm X^*} \bm  \beta  - \bm \Sigma_{\bm X^*} \bm  b - \bm \Omega_{\bm \eta}\widetilde{\bm \beta}  +  \bm \Sigma_{\bm \eta,\eps} +\bm \Omega_{\bm \eta,\eps}
=\bm \Sigma_{\bm X^*}  \bm \beta  +  \bm \Sigma_{\bm \eta,\eps} ,\\
\sigma_{Y}^2&=\widetilde{\bm \beta}'  {\bm \Sigma}_{\widetilde{\bm X}^*} \widetilde{\bm \beta} +\sigma_{\widetilde \eps}^2
= \widetilde{\bm \beta}' (\bm \Sigma_{\bm X^*} -\bm \Omega_{\bm \eta})  \widetilde{\bm \beta} +( \sigma_{\eps}^2+\Omega_{\eps})\\
&= \bm \beta' \bm \Sigma_{\bm X^*}  \bm \beta +\bm b'\bm \Sigma_{\bm X^*}  \bm b-2 \bm \beta '\bm \Sigma_{\bm X^*} \bm  b - \widetilde{\bm  \beta}' \bm \Omega_{\bm \eta}  \widetilde{\bm \beta} + \sigma_{\eps}^2+\Omega_{\eps}
= \bm \beta' \bm \Sigma_{\bm X^*} \bm  \beta +\sigma_{\eps}^2 .
\end{align*}
Choose $\bm \Omega_{\bm \eta,\eps}=\bm 0$. 
Then,   $\widetilde {\bm \Sigma} \succ 0$ and $\bm \Sigma_{\widetilde{\bm X}^*} \succ 0$ for small enough $||\bm b||$ 
because $\bm \Sigma \succ 0$, $\bm \Sigma_{{\bm X}^*} \succ 0$, and $||\bm \Omega|| \leq ||\bm b|| \cdot P_0$.
\end{enumerate}

\subsection{Proof of Theorem \protect\ref{th:eiv}} \label{ap:eiv}

Assume, without loss of generality, that all random variables have zero means. 
This simply avoids the extra term $i\sum_{k=1}^{p+1} a_{0k} s_{k}$ in some of the expressions.
Recall that we assume that 
$\beta_k \neq 0$, for $k=1,\ldots,p$, 
and that there does not exist a nonzero $\bm b$ such that $\bm b'\bm X^*$ is degenerate.
Otherwise there is multicollinearity and $\bm b'\bm X^*$ can be added to $\bm \beta'\bm X^*$ and subtracted from $\eps$ to obtain an observationally equivalent model even in the linear regression model without measurement errors
(i.e., $\bm \eta=\bm 0$).

Let 
$(\bm X^*,\bm \eta,\eps,\bm \beta)$ and $(\widetilde{\bm X}^*,\widetilde{\bm \eta},\widetilde \eps,\widetilde{\bm \beta})$
be observationally equivalent. Then,
\begin{align*}
\begin{pmatrix} \bm X \\ Y \end{pmatrix}
&=
\begin{pmatrix} \bm I_p \\ \bm \beta' \end{pmatrix}\bm X^*
+\begin{pmatrix} \bm \eta\\ \eps \end{pmatrix}
=
\begin{pmatrix} \bm I_p \\ \widetilde{\bm \beta}' \end{pmatrix}\widetilde{\bm X}^*
+\begin{pmatrix} \widetilde{\bm \eta} \\ \widetilde{\eps} \end{pmatrix}. 
\end{align*}

The log characteristic function of  $(\bm X,Y)$  is the functional equation,
\begin{align}
\varphi_{\bm X, Y}(\bm s )
&=\varphi_{\bm X^*}((\bm I_{p}  , \bm \beta)\bm s) 
+ \varphi_{ {\bm \eta}, \eps}(\bm s) 
=\varphi_{\widetilde{\bm X}^*}((\bm I_{p}  , \widetilde{\bm \beta})\bm s ) 
+ \varphi_{\widetilde{\bm \eta},\widetilde \eps}(\bm s),  &\bm s &\in \mathbb{R}^{p+1}, \label{eq:lcfY2}
\end{align}
where Assumption \ref{as:eivdep} that $\bm X^*$ is independent of $(\bm \eta,\eps)$ is used to separate log characteristic functions.
Hence,
\begin{align}
&\varphi_{\bm X^*}((\bm I_{p}  , \bm \beta)\bm s) 
-\varphi_{\widetilde{\bm X}^*}((\bm I_{p}  , \widetilde{\bm \beta})\bm s )  =  g(\bm s), \label{eq:eivlcf2}
\end{align}
where either by Assumption \ref{as:eivdep}(i) (independent errors), $g(\bm s) =\sum_{k=1}^{p+1}g_k(s_k)$, where 
$g_k(s_{k})=\varphi_{\widetilde \eta_k}( s_k)-\varphi_{ \eta_k}(s_k)$, for $k=1,\ldots,p$, and 
$g_{p+1}(s_{p+1})=\varphi_{\widetilde \eps} (s_{p+1}) -\varphi_{ \eps}(s_{p+1})$,
or by  Assumption \ref{as:eivdep}(ii) (joint normality of the errors), $g(\bm s)=-\frac{1}{2} \bm s'\bm \Omega \bm s$,
where  $\bm \Omega= \widetilde{\bm \Sigma}-\bm \Sigma$ is a $(p+1) \times (p+1)$ symmetric matrix.

\medskip

\begin{enumerate}[(i)]
	
\item \textbf{Sufficiency:} Assume that  $\bm \beta$ is not identified (i.e., $ \widetilde{\bm \beta}\neq \bm \beta$). 
We show that Assumptions \ref{as:eivdep}(i)-(v) hold.
The first derivatives of \eqref{eq:eivlcf2} evaluated at $\bm s=(\bm I_{p} , \bm 0)'\bm u + (-\widetilde{\bm\beta}',1)'  v=((\bm u-\widetilde{\bm\beta}v)',v)'$ are,
\begin{align*}
\left. \frac{\partial \varphi_{\bm X^*}(\bm w)}{\partial w_{k}}   \right|_{\bm w =(\bm u +(\bm \beta-\widetilde{\bm\beta})v)}
-  \left. \frac{\partial \varphi_{\widetilde{\bm X}^*}(\bm w)}{\partial w_{k}}   \right|_{\bm w =\bm u}& =g_{k}'(u_k-\widetilde \beta_k v), &k=&1,\ldots,p,\\
\sum_{k=1}^p \beta_{k} 
\left. \frac{\partial \varphi_{\bm X^*}(\bm w)}{\partial w_{k}}   \right|_{\bm w =(\bm u +(\bm \beta-\widetilde{\bm\beta})v)}
- \sum_{k=1}^p \widetilde \beta_{k} \left. \frac{\partial \varphi_{\widetilde{\bm X}^*}(\bm w)}{\partial w_{k}}   \right|_{\bm w =\bm u}&=g_{p+1}'(v).
\end{align*}
Next, the derivatives with respect to $v$ evaluated at $\bm u=-(\bm \beta-\widetilde{\bm\beta})v$ are,
\begin{align*}
\sum_{k_2=1}^p (\beta_{k_2}-\widetilde{\beta}_{k_2})
\left. \frac{\partial \varphi_{\bm X^*}^2(\bm w)}{\partial w_{k_1}\partial w_{k_2}}   \right|_{\bm w =\bm 0}&=-\widetilde \beta_{k_1} g_{k_1}''(- \beta_{k_1} v), &k_1=&1,\ldots,p,\\
\sum_{k_1,k_2} \beta_{k_1} (\beta_{k_2}-\widetilde{\beta}_{k_2})
\left. \frac{\partial \varphi_{\bm X^*}^2(\bm w)}{\partial w_{k_1}\partial w_{k_2}}   \right|_{\bm w =\bm 0} &=g_{p+1}''(v).
\end{align*}
Hence, $g_{k}''(s_k)=2\alpha_k$, where $\alpha_k=-\frac{1}{2\widetilde \beta_k}\sum_{k_2=1}^p (\beta_{k_2}-\widetilde{\beta}_{k_2})\cov(X^*_{k},X^*_{k_2})$, for $k=1,\ldots,p$, and $\alpha_{p+1}=\frac{1}{2}\sum_{k_1,k_2} \beta_{k_1} (\beta_{k_2}-\widetilde{\beta}_{k_2})\cov(X^*_{k_1},X^*_{k_2})$. Integrating the above equations we obtain $g_{k}(s_{k})=\alpha_ks_k^2$
in a neighborhood of the origin.\footnote{\cite{khatri1972functional} show that $g_{k}(s_{k})=\alpha_ks_k^2$ in a neighborhood of the origin 
	using finite differences (and continuity of the log characteristic function) instead of derivatives so no moments need to exist up to this point in the proof.} By analytic continuation $g_{k}(s_{k})=\alpha_ks_k^2$, $s_k \in \mathbb{R}$, for $k=1,\ldots,p+1$. 
Substituting back into \eqref{eq:eivlcf2} we obtain,
\begin{align}
\varphi_{\bm X^*}((\bm I_{p}  , \bm \beta)\bm s) 
-\varphi_{\widetilde{\bm X}^*}((\bm I_{p}  , \widetilde{\bm \beta})\bm s )  =  \sum_{k=1}^{p+1} g_k(s_k) =   \sum_{k=1}^{p+1} \alpha_{k}s_k^2. \label{eq:eivlcf21}
\end{align}	

			Next, substitute $\bm s=((\bm u -\widetilde{\bm\beta}v)',v)' $,  $\bm s= (-\widetilde{\bm\beta}',1)'  v$, 
			and $\bm s=(\bm u, 0)'$ into \eqref{eq:eivlcf21} respectively,
			\begin{align}
				\varphi_{\bm X^*}(\bm u +\bm b  v) - \varphi_{\widetilde{\bm X}^*}(\bm u ) 
				&= \sum_{k=1}^{p} g_{k}(u_k-\widetilde \beta_k v) +g_{p+1}(v)
				= \sum_{k=1}^{p} \alpha_k (u_k-\widetilde \beta_k v)^2 +\alpha_{p+1}v^2, \label{eq:eivind1}\\
				\varphi_{\bm X^*}(\bm b  v) 
				&= \sum_{k=1}^{p} g_{k}(\widetilde \beta_k v) +g_{p+1}(v)
				= \sum_{k=1}^{p} \alpha_k \widetilde \beta_k^2 v^2 +\alpha_{p+1}v^2, \label{eq:eivind2} \\
				\varphi_{\bm X^*}(\bm u) - \varphi_{\widetilde{\bm X}^*}(\bm u ) 
				&= \sum_{k=1}^{p} g_{k}(u_k) = \sum_{k=1}^{p} \alpha_k u_k^2, \label{eq:eivind3}
			\end{align}	
where $\bm b=\bm \beta- \widetilde{\bm \beta}$. 
By \eqref{eq:eivind2} and \cite{marcinkiewicz1939propriete}, $\bm b'\bm X^*$ is normal.\footnote{Even without analytic continuity, \eqref{eq:eivind2} holds in a neighborhood of the origin so $\bm b'\bm X^*$ is normal by \cite{marcinkiewicz1939propriete}.}
If there does not exist a nonzero $\bm b\in \mathbb{R}^p$ such that $\bm b'\bm X^*$ is normal then $\bm b=\bm 0$ and $\bm \beta$ is identified. 
Hence, nonidentifiability 
implies that $\bm b' \bm X^*$  
is nondegenerate normal (Assumption \ref{as:eiv}(i) holds).

Substitute \eqref{eq:eivind2} and \eqref{eq:eivind3} into \eqref{eq:eivind1} to obtain,
			\begin{align}
			\varphi_{\bm X^*}(\bm u +\bm b  v) 
			&=\varphi_{ {\bm X^*}}(\bm u ) 	+\varphi_{\bm X^*}(\bm b  v) 
			-2v \bm d' \bm u, 
			\label{eq:eivB}
			\end{align}
where $\bm d=\begin{pmatrix}
\alpha_1 \widetilde \beta_1 & \ldots &  \alpha_p \widetilde \beta_p
\end{pmatrix}'$.
Next, let $\bm B\in \mathbb{R}^{p\times(p-1)}$ 
have full column rank and satisfy
\begin{align}
\bm d'\bm B =\bm 0, \label{eq:orth}
\end{align} 
and substitute $\bm u=\bm B \bm w$, for $\bm w\in \mathbb{R}^{p-1}$, into \eqref{eq:eivB}, 
to obtain,	
			\begin{align}
			\varphi_{\bm X^*}(\bm B \bm w +\bm b  v) 
			&=\varphi_{ {\bm X^*}}(\bm B \bm w) 	+\varphi_{\bm X^*}(\bm b  v) . \label{eq:eivindep}
			\end{align}
Hence, nonidentifiability implies that $\bm B' \bm X^*$  and $\bm b' \bm X^*$ are independent (Assumption \ref{as:eiv}(ii) holds).
Now, if $(\bm B , \bm b)$ is singular then some nonzero linear combination of the columns of $\bm B$ is equal to $\bm b$, 
which implies that $\bm b'\bm X^*$ is independent of itself and $\bm b'\bm X^*$ is degenerate.
Hence, nonidentifiability 
implies that  $(\bm B , \bm b)$ is nonsingular (Assumption \ref{as:eiv}(iii) holds).

Next substitute $\bm s=((\bm B\bm w+\bm b z-\widetilde{\bm \beta}v)',v)'$ into \eqref{eq:eivlcf2} and simplify to obtain,
\begin{align*}
0&= (\sum_{k=1}^{p} d_k\widetilde \beta_k    + \frac{1}{2}\bm b' \bm \Sigma_{\bm X^*}\bm b+ \alpha_{p+1}) v^2
+(\bm b' \bm \Sigma_{\bm X^*}\bm b -2\sum_{k=1}^{p} d_kb_k)vz
-2\bm d' \bm B\bm w  v.
\end{align*} 
Using \eqref{eq:orth} and equating coefficients we obtain,
\begin{align*}
\frac{1}{2}\bm b'\bm  \Sigma_{\bm X^*} \bm b&=\sum_{k=1}^pd_kb_k 
=-\sum_{k=1}^pd_k\widetilde \beta_k-\alpha_{p+1}>0 .
\end{align*}
By the second equality $\alpha_{p+1}=-\sum_{k=1}^pd_k\beta_k$. 
By the first equality and \eqref{eq:orth}, $\bm d$ solves the $p$ linear equations,
\begin{align*}
\bm d'
\begin{pmatrix}
\bm B &
\bm b
\end{pmatrix}
=
\begin{pmatrix}
\bm 0 &
\frac{1}{2}\bm b'\bm  \Sigma_{\bm X^*} \bm b
\end{pmatrix}.
\end{align*}
But by Assumption \ref{as:eiv}(ii), 
$\bm 0 = \cov(\bm B' \bm X^*,\bm b'\bm X^*)=\bm B'\cov(\bm X^*,\bm b'\bm X^*)$.
Noting that $\bm b'\bm  \Sigma_{\bm X^*} \bm b = \bm b'\cov(\bm X^*,\bm b'\bm X^*)$,
we obtain the solution $\bm d=\frac{1}{2}\cov(\bm X^*,\bm b'\bm X^*)$. 
By Assumption \ref{as:eiv}(iii), $(\bm B, \bm b)$ is nonsingular so that
this is the unique solution.
Hence, $\alpha_k= d_k/\widetilde \beta_k= \frac{1}{2}\cov(X^*_k,\bm b' \bm X^*)/\widetilde \beta_k$, for $k=1,\ldots,p$,
and $\alpha_{p+1}=-\sum_{k=1}^pd_k\beta_k= -\frac{1}{2}\cov(\bm \beta' \bm X^*,\bm b'\bm X^*)$.

Recall that $g_k(s_{k})=\varphi_{\widetilde \eta_k}( s_k)-\varphi_{\eta_k}(s_k)=\alpha_ks_k^2$, for $k=1,\ldots,p+1$, so
if $-2\alpha_{p+1}=2\sum_{k=1}^pd_k\beta_k= \cov(\bm \beta '\bm X^*,\bm b'\bm X^*)<0$ then $\eps$ is divisible by a nondegenerate normal distribution (Assumption \ref{as:eiv}(iv) holds) and
if $2\alpha_k= 2d_k/\widetilde \beta_k= \cov(X^*_k,\bm b' \bm X^*)/\widetilde \beta_k> 0$ then $\eta_k$ is divisible by a nondegenerate normal distribution (Assumption \ref{as:eiv}(v) holds).

Finally, $ \sigma_{\widetilde \eps}^2 =\sigma_{\eps}^2+\sum_{k=1}^p d_k \beta_k$, $\sigma_{\widetilde \eta_k}^2=  \sigma_{ \eta_k}^2 -\frac{d_k}{\widetilde \beta_k}$, 
$\sigma_{{\widetilde X}^*_k}^2= \sigma_{X^*_k}^2 +\frac{d_k}{\widetilde \beta_k} $, and $ \sigma_{{\widetilde X}^*_k{\widetilde X}^*_j}=\sigma_{X^*_kX^*_j}$.
Hence, $||\bm b||$ needs to be small enough so that 
$ \sigma_{\widetilde \eps}^2>0$,  $\sigma_{\widetilde \eta_k}^2>0$, and $\bm \Sigma_{\widetilde{\bm X}^*}\succ 0$, which follows  
by $ \sigma_{\eps}^2>0$, $\sigma_{ \eta_k}^2>0$, $\bm \Sigma_{{\bm X}^*} \succ 0$, and $|d_k|\leq ||\bm b||\cdot ||\bm \Sigma_{{\bm X}^*}  ||$.

\medskip

\textbf{Necessity:}
Assume that Assumptions \ref{as:eiv}(i)-(v) hold. We show that $\bm \beta$ is not identified by constructing an observationally equivalent model that satisfies \eqref{eq:lcfY2}.
By Assumptions \ref{as:eiv}(i)-(iii) there exists a nonsingular matrix $(\bm B , \bm b)$ such that $\bm B'\bm X^*$ and $\bm b'\bm X^*\sim N(0,\bm b'\bm  \Sigma_{\bm X^*} \bm b)$ are independent with $\bm b'\bm  \Sigma_{\bm X^*} \bm b>0$ and $||\bm b||$ small. 
Let $Z_{1k}\sim N(0,|{\cov(X^*_k,\bm b'\bm X^*)}/{\widetilde \beta_k}|)$, $Z'_{1k}\sim N(0,|{\cov(X^*_k,\bm b'\bm X^*)}/{\widetilde \beta_k}|)$, and $Z_2\sim N(0,|\cov(\bm \beta' \bm X^*,\bm b'\bm X^*)|)$
be independent random variables.
Define $\widetilde{\bm \beta}=\bm \beta- \bm b$.
If $\cov(X^*_k,\bm b'\bm X^*)/\widetilde \beta_k>0$ then define ${\widetilde X}^*_k= X^*_k +Z_{1k}$ and $ \eta_k= \widetilde \eta_k +Z'_{1k}$ otherwise define $ X^*_k= {\widetilde X}^*_k +Z_{1k}$ and $\widetilde \eta_k =   \eta_k +Z'_{1k}$, and if $\cov(\bm \beta' \bm X^*,\bm b'\bm X^*)<0$ then define $ \eps =\widetilde \eps + Z_2$ otherwise define $\widetilde \eps =\eps + Z_2$.			
		
		Then $\varphi_{\widetilde{\bm X}^*}(\bm u ) =\varphi_{\bm X^*}(\bm u) - \sum_{k=1}^{p} \alpha_k u_k^2$, 
		$\varphi_{\widetilde{\eta}_k}(s_k )=\varphi_{\eta_k}(s_k) + \alpha_ks_k^2$, 
		and $\varphi_{\widetilde{\eps}}(s_{p+1} )= \varphi_{\eps}(s_{p+1})+\alpha_{p+1} s_{p+1}^2 $,				
		where $\alpha_k=\frac{1}{2}\cov(X^*_k,\bm b'\bm X^*)/\widetilde \beta_k$ and $\alpha_{p+1}=-\frac{1}{2}\cov(\bm \beta'\bm X^*,\bm b'\bm X^*)$.
				 
		Substitute $\bm s=((\bm I_p , \bm 0)'(\bm B , \bm b),(-\widetilde{\bm \beta}',1)')(\bm w' , z,v)'
		=((\bm B\bm w+\bm b z-\widetilde{\bm \beta}v)',v)'\in \mathbb{R}^{p+1}$ 
		into \eqref{eq:lcfY2} to obtain,
		{\small	\begin{align*}
			\varphi_{\bm X, Y}(\bm s)
			&=\varphi_{ \widetilde{\bm X}^*}((\bm I_{p} , \widetilde{\bm \beta})\bm s) + \sum_{k=1}^p\varphi_{\widetilde \eta_k}(s_k)+\varphi_{\widetilde \eps}(s_{p+1})\\
			&=\varphi_{ \widetilde{\bm X}^*}(\bm B\bm w +\bm b z) 
		+\sum_{k=1}^{p} \varphi_{\widetilde \eta_k}(\bm B'_k \bm w+b_kz-\widetilde \beta_kv)
		+ \varphi_{ \widetilde \eps}(v)\\
			&=\left[\varphi_{ {\bm X^*}}(\bm B\bm w +\bm b z) 
		-\sum_{k=1}^{p} \alpha_k(B_k'\bm w +b_k z)^2\right]\\
		& +\sum_{k=1}^{p} \left[\varphi_{ \eta_k}(\bm B'_k \bm w+b_kz-\widetilde \beta_kv)
		+ \alpha_k(\bm B'_k \bm w+b_kz-\widetilde \beta_kv)^2\right]
		+ \left[\varphi_{ \eps}(v)+\alpha_{p+1}v^2\right]\\
			&=\varphi_{ {\bm X^*}}(\bm B\bm w)
		+\sum_{k=1}^{p} \varphi_{ \eta_k}(\bm B'_k \bm w+b_kz-\widetilde \beta_kv)
		+ \varphi_{ \eps}(v)-\frac{1}{2}\bm b'\Sigma_{\bm X^*}\bm b z^2\\
		&
		-2\sum_{k=1}^{p} \alpha_k \widetilde \beta_k\bm B'_k \bm w v
		-2\sum_{k=1}^{p} \alpha_k\widetilde \beta_kb_kzv
		+\sum_{k=1}^{p} \alpha_k\widetilde \beta_k^2v^2
		+\alpha_{p+1}v^2\\
			&=\varphi_{ {\bm X^*}}(\bm B\bm w)
			+\sum_{k=1}^{p} \varphi_{ \eta_k}(\bm B'_k \bm w+b_kz-\widetilde \beta_kv)
			+ \varphi_{ \eps}(v)-\frac{1}{2}\bm b'\Sigma_{\bm X^*}\bm b z^2\\
			&	-\cov(\bm B'\bm X^*,\bm b'\bm X^*) \bm w v
			-\cov(\bm b'\bm X^*,\bm b'\bm X^*) zv
			+\frac{1}{2}\cov(\widetilde{\bm\beta}'\bm X^*,\bm b'\bm X^*)v^2
			-\frac{1}{2}\cov(\bm \beta'\bm X^*,\bm b'\bm X^*)v^2\\
			&=\varphi_{ {\bm X^*}}(\bm B\bm w)
	+\sum_{k=1}^{p} \varphi_{ \eta_k}(\bm B'_k \bm w+b_kz-\widetilde \beta_kv)
	+ \varphi_{ \eps}(v)-\frac{1}{2}\bm b'\Sigma_{\bm X^*}\bm b (v+z)^2\\
		&=\varphi_{ {\bm X^*}}((\bm I_{p} , { \bm \beta})\bm s) +\sum_{k=1}^{p}\varphi_{ \eta_k}(s_{p})+ \varphi_{  \eps}(s_{p+1}),
			\end{align*}
		}where $\bm B_k= (B_{k1},\ldots,B_{kp-1})'$ is the $k$-th row of $\bm B$. 
	The fourth, sixth, and last equalities follow by Assumptions \ref{as:eiv}(i)-(ii) that $\bm B'\bm X^*$ and $\bm b'\bm X^*$ are independent (so $\cov(\bm B'\bm X^*,\bm b'\bm X^*)=0$) and $\bm b'\bm X^*$ is normal.
	
\medskip
	
\item \textbf{Sufficiency:} Assume that $\bm \beta$ is not identified. We show that Assumptions \ref{as:eiv}(i)-(iii) hold.
		Let
		\begin{align}
		\bm \Omega=	\begin{pmatrix}  \bm \Omega_{\bm  \eta}& \bm \Omega_{\bm  \eta,\eps}\\
		\bm \Omega_{\bm  \eta,\eps}' &  \Omega_{\eps}
		\end{pmatrix}, \quad 
		\widetilde{\bm \Sigma}=  \begin{pmatrix}  {\bm \Sigma}_{\widetilde{\bm \eta}}& {\bm \Sigma}_{\widetilde{\bm \eta},\widetilde\eps} \\
		{\bm \Sigma}_{\widetilde{\bm \eta},\widetilde\eps}'  & \sigma^2_{\widetilde\eps}
		\end{pmatrix}, \quad
		\bm \Sigma = \begin{pmatrix}\bm \Sigma_{\bm \eta}  & \bm \Sigma_{\bm \eta,\eps}\\
		\bm \Sigma_{\bm \eta,\eps}' & \sigma^2_{\eps},
		\end{pmatrix} \label{eq:sig}
		\end{align}
		and substitute 
		 $\bm s=((\bm u -\widetilde{\bm\beta}v)',v)' $,  $\bm s= (-\widetilde{\bm\beta}',1)'  v$, and $\bm s=(\bm u, 0)'$ into \eqref{eq:eivlcf2} respectively,
		\begin{align}
		\varphi_{\bm X^*}(\bm u +\bm b  v) 
		&=\varphi_{\widetilde{\bm X}^*}(\bm u ) 
		-\frac{1}{2} \bm u' {\bm \Omega}_{\bm \eta}\bm u
		+v(\widetilde{\bm\beta}' {\bm \Omega}_{\bm \eta}- {\bm \Omega}_{\bm \eta,\eps}')\bm u 
		-\frac{1}{2}(\widetilde{\bm\beta}' {\bm \Omega}_{\bm \eta}\widetilde{\bm\beta}-2 {\bm \Omega}_{\bm \eta,\eps}'\widetilde{\bm\beta}+  {\Omega}_{\eps})v^2,\label{eq:eivN1} \\
		\varphi_{\bm X^*}(\bm b  v) 	&
		= -\frac{1}{2}(\widetilde{\bm\beta}'{\bm \Omega}_{\bm \eta}\widetilde{\bm\beta}
		-2{\bm \Omega}_{\bm \eta,\eps}'\widetilde{\bm\beta}+{ \Omega}_{\eps})v^2
		, \label{eq:eivn2} \\
		\varphi_{\bm X^*}(\bm u) 	&=\varphi_{\widetilde{\bm X}^*}(\bm u ) 
		-\frac{1}{2} \bm u'  {\bm \Omega}_{\bm \eta}\bm u, \label{eq:eivn3}
		\end{align}	
		where $\bm b =\bm \beta- \widetilde{\bm \beta}$. 
		By \eqref{eq:eivn2} and \cite{marcinkiewicz1939propriete}, $\bm b'\bm X^*$ is normal.
		If there does not exist a nonzero $\bm b\in \mathbb{R}^p$ such that $\bm b'\bm X^*$ is normal then $\bm b=\bm 0$ and $\bm \beta$ is identified. 
		Hence, nonidentifiability implies that $\bm b' \bm X^*$ 
		is nondegenerate normal (Assumption \ref{as:eiv}(i) holds).

		Substitute \eqref{eq:eivn2} and \eqref{eq:eivn3} into \eqref{eq:eivN1} to obtain,
		\begin{align}
		\varphi_{\bm X^*}(\bm u +\bm b  v) 
		&=\varphi_{\bm X^*}(\bm u)+ \varphi_{\bm X^*}(\bm b  v)+ v(\widetilde{\bm\beta}'\bm \Omega_{\bm \eta}
		-\bm \Omega_{\bm \eta,\eps}')\bm u . \label{eq:eivn1}
		\end{align}		
		Next, let $\bm B\in \mathbb{R}^{p\times(p-1)}$
		have full column rank and satisfy
		\begin{align}
		(\widetilde{\bm\beta}'\bm \Omega_{\bm \eta}
		-\bm \Omega_{\bm \eta,\eps}')\bm B =\bm 0, \label{eq:eivn}
		\end{align}  
		and substitute $\bm u=\bm B \bm w$, for $\bm w\in \mathbb{R}^{p-1}$, into \eqref{eq:eivn1} to obtain,
		\begin{align}
		\varphi_{\bm X^*}(\bm B \bm w +\bm b  v) 
		&=\varphi_{ {\bm X^*}}(\bm B \bm w) 	+\varphi_{\bm X^*}(\bm b  v). \label{eq:eivnindep}
		\end{align}
		Hence, nonidentifiability implies that $\bm B' \bm X^*$ and $\bm b' \bm X^*$ are independent  (Assumption \ref{as:eiv}(ii) holds). 
		Now, if $(\bm B , \bm b)$ is singular then some nonzero linear combination of the columns of $\bm B$ is equal to $\bm b$, which implies that $\bm b'\bm X^*$ is independent of itself and $\bm b'\bm X^*$ is degenerate. 
		Hence, nonidentifiability 
		implies that  $(\bm B , \bm b)$ is nonsingular  (Assumption \ref{as:eiv}(iii) holds).

Next substitute $\bm s=((\bm B\bm w+\bm b z-\widetilde{\bm \beta}v)',v)'$ into \eqref{eq:eivlcf2} and simplify to obtain,
{\small 
\begin{align*}
0&= (\widetilde{\bm \beta}'  {\bm \Omega}_{\bm \eta}-{\bm \Omega}_{\bm \eta,\eps}')\bm B\bm wv
-({\bm \Omega}_{\bm \eta,\eps}'\bm b- \widetilde{\bm \beta}' {\bm \Omega}_{\bm \eta}\bm b-\bm b'\bm  \Sigma_{\bm X^*} \bm b) zv -\frac{1}{2} (\widetilde{\bm \beta}' {\bm \Omega}_{\bm \eta} \widetilde{\bm \beta}-2 {\bm \Omega}_{\bm \eta,\eps}'\widetilde{\bm \beta}+ \Omega_{\eps}-\bm b'\bm  \Sigma_{\bm X^*} \bm b) v^2.
\end{align*}	
}Using \eqref{eq:eivn} and equating coefficients we obtain,	
\begin{align*}
\begin{pmatrix}
{\bm \Omega}_{\bm \eta,\eps}' - \widetilde{\bm \beta}'{\bm \Omega}_{\bm \eta}, &
\Omega_{\eps}- {\bm \Omega}_{\bm \eta,\eps}'\widetilde{\bm \beta}
\end{pmatrix}
&=\begin{pmatrix}
(\bm 0 ,\bm b' \bm \Sigma_{\bm X^*} \bm b)(\bm B,\bm b)^{-1}, &
(\bm 0 ,\bm b' \bm \Sigma_{\bm X^*} \bm b)(\bm B,\bm b)^{-1}\widetilde{\bm \beta}+\bm b' \bm \Sigma_{\bm X^*} \bm b
\end{pmatrix}\\
&=\begin{pmatrix}
\bm b'\bm \Sigma_{\bm X^*} ,&
\bm b'\bm \Sigma_{\bm X^*}\bm \beta
\end{pmatrix},
\end{align*}
where the last equality follows because
  \begin{align*}
    (\bm 0 ,\bm b' \bm \Sigma_{\bm X^*} \bm b)(\bm B,\bm b)^{-1}
  =(\bm b'\bm \Sigma_{\bm X^*}\bm B ,\bm b' \bm \Sigma_{\bm X^*} \bm b)(\bm B,\bm b)^{-1}
  =\bm b'\bm \Sigma_{\bm X^*}(\bm B , \bm b)(\bm B,\bm b)^{-1}=\bm b'\bm \Sigma_{\bm X^*},
  \end{align*}
  where $\bm 0=\cov(\bm b'\bm X^*,\bm B'\bm X^*) =\bm b' \bm \Sigma_{\bm X^*} \bm B$ by Assumption \ref{as:eiv}(ii) and
  $(\bm B,\bm b)$ is nonsingular by Assumption \ref{as:eiv}(iii).  
Hence,  $ \widetilde{\bm \beta}'{\bm \Omega}_{\bm \eta}={\bm \Omega}_{\bm \eta,\eps}' -\bm b'\bm \Sigma_{\bm X^*}$ and $\Omega_{\eps}=\bm b'\bm \Sigma_{\bm X^*}{\bm \beta}
+ {\bm \Omega}_{\bm \eta,\eps}'\widetilde{\bm \beta}$. Let ${\bm \Omega}_{\bm \eta,\eps}=\bm 0$ and solve for $\bm \Omega$ to obtain,
\begin{align}
\bm \Omega&=
\begin{pmatrix}
- \cov(\frac{X^*_1}{\widetilde \beta_1},\bm b'\bm X^*) &0 &\cdots &0\\
0& \ddots & \ddots& \vdots \\
 \vdots & \ddots &- \cov(\frac{X^*_p}{\widetilde \beta_p},\bm b'\bm X^*) &0\\
  0 & \cdots & 0 &\cov( \bm \beta '\bm X^*, \bm b'\bm X^*)
\end{pmatrix}. \label{eq:Omega}
\end{align}
Finally, 
$||\bm b||$ needs to be small enough so that  $\widetilde {\bm \Sigma} \succ 0$ and $\bm \Sigma_{\widetilde{\bm X}^*} \succ 0$, which follows by 
$\bm \Sigma \succ 0$, $\bm \Sigma_{{\bm X}^*} \succ 0$, and $||\bm \Omega|| \leq ||\bm b|| \cdot P_0$.

\medskip

\textbf{Necessity:}	Assume that Assumptions \ref{as:eiv}(i)-(iii) hold. We construct an observationally equivalent model that satisfies \eqref{eq:lcfY2}.
By Assumptions \ref{as:eiv}(i)-(iii), there exists a nonsingular matrix $(\bm B , \bm b)$
		such that  $\bm B'\bm X^*$ and $\bm b'\bm X^* \sim N(0,\bm b'\bm  \Sigma_{\bm X^*} \bm b)$ are independent with $\bm b'\bm  \Sigma_{\bm X^*} \bm b>0$ and $||\bm b||$ small.
		Let $Z_{1k}\sim N(0,|{\cov(X^*_k,\bm b'\bm X^*)}/{\widetilde \beta_k}|)$ 
		be an independent random variable.
		Define $\widetilde{\bm \beta}=\bm \beta- \bm b$,
		$(\widetilde{\bm \eta},\widetilde \eps)\sim N(\bm 0,\widetilde{\bm \Sigma})$, where $\widetilde{\bm \Sigma}=\bm \Sigma +\bm \Omega$ with  $\bm \Sigma$ and $\bm \Omega$ defined in 
		\eqref{eq:sig} and \eqref{eq:Omega},
		and if $\cov(X^*_k,\bm b'\bm X^*)/\widetilde \beta_k>0$ then ${\widetilde X}^*_k= X^*_k +Z_{1k}$ 
		otherwise $ X^*_k= {\widetilde X}^*_k +Z_{1k}$. 
		
		Then $\varphi_{ \widetilde{\bm X^*}}(\bm u)=\varphi_{{ \bm X^*}}(\bm u)+\frac{1}{2}\bm u' \bm \Omega_{\bm \eta} \bm u$, 
		and $\varphi_{\widetilde{\bm \eta},\widetilde \eps}(\bm s)= -\frac{1}{2}\bm s'  \widetilde{\bm \Sigma}\bm s $. 	
		Substitute $\bm s=((\bm B\bm w+\bm b z-\widetilde{\bm \beta}v)',v)'\in \mathbb{R}^{p+1}$
		into \eqref{eq:lcfY2} to obtain,
		{\small\begin{align*} 
		\varphi_{\bm X,Y}(\bm s)&=\varphi_{\widetilde{\bm X}^*}((\bm I_{p}  , \widetilde{\bm \beta})\bm s ) 
		-\frac{1}{2} \bm s' \widetilde{\bm \Sigma} \bm s\\
		&=\varphi_{ \widetilde{\bm X^*}}(\bm B\bm w +\bm bz)
		-\frac{1}{2}((\bm B\bm w+\bm bz-\widetilde{\bm \beta}v)',v)\begin{pmatrix}  {\bm \Sigma}_{\widetilde{\bm \eta}}& {\bm \Sigma}_{\widetilde{\bm \eta},\widetilde\eps} \\
		{\bm \Sigma}_{\widetilde{\bm \eta},\widetilde\eps}'  &  \sigma^2_{\widetilde\eps} 
		\end{pmatrix}\begin{pmatrix}
		\bm B\bm w+\bm bz-\widetilde{\bm \beta}v\\v
		\end{pmatrix}\\
		&=\left[\varphi_{  \bm X^*}(\bm B\bm w +\bm bz)+\frac{1}{2}(\bm B\bm w+\bm bz)'\bm  \Omega_{\bm \eta} (\bm B\bm w+\bm bz)\right]\\
		& -\frac{1}{2}((\bm B\bm w+\bm bz-\widetilde{\bm \beta}v)',v)\left[\begin{pmatrix} {\bm \Sigma}_{\bm \eta}& {\bm \Sigma}_{\bm \eta,\eps} \\
		{\bm \Sigma}_{\bm \eta,\eps}'  &  \sigma^2_{\eps} 
		\end{pmatrix}
		+
		\begin{pmatrix} {\bm \Omega}_{\bm \eta}& {\bm \Omega}_{\bm \eta,\eps} \\
		{\bm \Omega}_{\bm \eta,\eps}'  &  \Omega_{\eps}
		\end{pmatrix}
		\right]
		\begin{pmatrix}
		\bm B\bm w+\bm bz-\widetilde{\bm \beta}v\\v
		\end{pmatrix}	\\
		&=\left[\varphi_{  \bm X^*}(\bm B\bm w)-\frac{1}{2}\bm b'\bm  \Sigma_{\bm X^*} \bm bz^2\right]
		-\frac{1}{2}((\bm B\bm w+\bm bz-\widetilde{\bm \beta}v)',v)\begin{pmatrix} {\bm \Sigma}_{\bm \eta}& {\bm \Sigma}_{\bm \eta,\eps} \\
		{\bm \Sigma}_{\bm \eta,\eps}'  &  \sigma^2_{\eps} 
		\end{pmatrix}\begin{pmatrix}
		\bm B\bm w+\bm bz-\widetilde{\bm \beta}v\\v
		\end{pmatrix}\\
		& +\frac{1}{2}(\bm B\bm w+\bm bz)'\bm  \Omega_{\bm \eta} (\bm B\bm w+\bm bz)	
		- \frac{1}{2}(\bm B\bm w+\bm bz-\widetilde{\bm \beta}v)' {\bm \Omega}_{\bm \eta}(\bm B\bm w+\bm bz-\widetilde{\bm \beta}v)\\
		&
		- {\bm \Omega}_{\bm \eta,\eps}'(\bm B\bm w+\bm bz-\widetilde{\bm \beta}v)v  - \frac{1}{2}\Omega_{\eps} v^2\\
		&=\varphi_{  \bm X^*}(\bm B\bm w)-\frac{1}{2}((\bm B\bm w+\bm bz-\widetilde{\bm \beta}v)',v)\begin{pmatrix} {\bm \Sigma}_{\bm \eta}& {\bm \Sigma}_{\bm \eta,\eps} \\
		{\bm \Sigma}_{\bm \eta,\eps}'  &  \sigma^2_{\eps} 
		\end{pmatrix}\begin{pmatrix}
		\bm B\bm w+\bm bz-\widetilde{\bm \beta}v\\v
		\end{pmatrix}\\
		&-\frac{1}{2}\bm b'\bm  \Sigma_{\bm X^*} \bm bz^2 -( {\bm \Omega}_{\bm \eta,\eps}'
		-\widetilde{\bm \beta}'  {\bm \Omega}_{\bm \eta})\bm B\bm wv
		- ({\bm \Omega}_{\bm \eta,\eps}'
		- \widetilde{\bm \beta}' {\bm \Omega}_{\bm \eta})\bm b zv
		- \frac{1}{2}(\widetilde{\bm \beta}' {\bm \Omega}_{\bm \eta} \widetilde{\bm \beta}
		-2 {\bm \Omega}_{\bm \eta,\eps}'\widetilde{\bm \beta}
		+ \Omega_{\eps}) v^2\\
		&=\varphi_{  \bm X^*}(\bm B\bm w)-\frac{1}{2}\bm b'\bm  \Sigma_{\bm X^*} \bm b(v+z)^2-\frac{1}{2}((\bm B\bm w+\bm bz-\widetilde{\bm \beta}v)',v)\begin{pmatrix} {\bm \Sigma}_{\bm \eta}& {\bm \Sigma}_{\bm \eta,\eps} \\
		{\bm \Sigma}_{\bm \eta,\eps}'  &  \sigma^2_{\eps} 
		\end{pmatrix}\begin{pmatrix}
		\bm B\bm w+\bm bz-\widetilde{\bm \beta}v\\v
		\end{pmatrix}\\
		&=\varphi_{  \bm X^*}((\bm I_p , {\bm \beta})\bm s)-\frac{1}{2}\bm s'{\bm \Sigma}\bm s,
		\end{align*}
	}where the fourth, sixth, and last equalities follow by Assumptions \ref{as:eiv}(i)-(ii) that $\bm B'\bm X^*$ and $\bm b'\bm X^*$ are independent (so 
$\bm b'\bm \Sigma_{X^*}\bm B=\cov(\bm b'\bm X^*,\bm B'\bm X^*)=0$) and $\bm b'\bm X^*$ is normal.
 
\end{enumerate}

\subsection{Proof of Corollary \protect\ref{co:simeiv}} \label{ap:co:simeiv}

	In the case $p=1$, \eqref{eq:eivlcf2} is,
	\begin{align}
&\varphi_{ X^*}(s_1 +  \beta s_2) 
-\varphi_{\widetilde{X}^*}(s_1 + \widetilde \beta s_2 )  =  g_1(s_1) + g_2(s_2), \label{eq:eivlcf2s}
\end{align}
where $g_1(s_{1})=\varphi_{\widetilde \eta}( s_1)-\varphi_{\eta}(s_1)$ and 
$g_{2}(s_{2})=\varphi_{\widetilde \eps} (s_{2}) -\varphi_{ \eps}(s_{2})$. 
By \cite{khatri1972functional}, $g_1(s_{1})=\alpha_{1}s_1^2$ and
$g_2(s_{2})=\alpha_{2}s_2^2$ in some neighborhood of the origin.

\medskip

\textbf{Sufficiency:} Assume that $\beta$ is not identified. 
We show that $X^*$ is normal and either
$\eps$ or $\eta$ is divisible by a nondegenerate normal distribution.
Substitute  $\bm s= (-\widetilde{\beta}v,v)'  $ and  $\bm s= (-{\beta}v,v)'  $
respectively into \eqref{eq:eivlcf2s} to obtain,
\begin{align}
\varphi_{X^*}( b  v) 
&=  g_{1}(-\widetilde \beta v) +g_{2}(v)
= \alpha_1 \widetilde \beta^2 v^2 +\alpha_{2}v^2, \label{eq:eivind1s} \\
-\varphi_{{\widetilde X}^*}( -b  v) 
&=  g_{1}(- \beta v) +g_{2}(v)
= \alpha_1  \beta^2 v^2 +\alpha_{2}v^2, \label{eq:eivind2s} 
\end{align}	 
where $b=\beta-\widetilde \beta$.
By 
\cite{marcinkiewicz1939propriete}, $bX^*$ and $b {\widetilde X}^*$ are normal. 
If $X^*$ is nongaussian then $b=0$ and $\beta$ is identified.
Hence, nonidentification implies that $X^*$ and ${\widetilde X}^*$ are nondegenerate normal.

Using \eqref{eq:eivlcf2s} and that $X^*$ and ${\widetilde X}^*$ are nondegenerate normal,
	\begin{align*}
	\frac{1}{2}(\sigma_{{\widetilde X}^*}^2-\sigma_{X^*}^2)s_1^2
	 +(\sigma_{{\widetilde X}^*}^2\widetilde \beta  -\sigma_{X^*}^2\beta )s_1s_2 
	 +\frac{1}{2}(\sigma_{{\widetilde X}^*}^2 \widetilde \beta^2  -\sigma_{X^*}^2 \beta^2 )s_2^2 
	&=  g_1(s_1) + g_2(s_2), &(s_1,s_2)\in \mathbb{R}^2.
\end{align*}
Equating coefficients, $\sigma_{{{\widetilde X}}^*}^2 \widetilde \beta  =\sigma_{X^*}^2 \beta$ ($|b|=|\beta-\widetilde \beta|$ needs to be small enough so that $\beta$ and $\widetilde \beta$ have the same sign),  
$g_1(s_1)=\alpha_1s_1^2$, and $g_2(s_2)=\alpha_2s_2^2$, where $\alpha_1=\frac{1}{2}\sigma_{X^*}^2b/\widetilde \beta$ and $\alpha_2=-\frac{1}{2}\sigma_{X^*}^2\beta b$. 
If $\alpha_1=\alpha_2=0$ then $b=0$ and $\beta$ is identified. 
Hence, nonidentifiability implies that either $\eps$ or $\eta$ is divisible by a nondegenerate normal distribution. 

\medskip

\textbf{Necessity:} 
Assume that $X^* \sim N(0,\sigma_{X^*}^2)$ with $\sigma_{X^*}^2>0$ and either $\eps$ or $\eta$ is divisible by a nondegenerate normal distribution.
We construct an observationally equivalent model that satisfies \eqref{eq:lcfY2}.
Define $\widetilde \beta=\beta-b$ for some small nonzero $b\in \mathbb{R}$ so that $\beta$ and $\widetilde \beta$ have the same sign,
and let  $Z_1 \sim N(0,|{b}/{\widetilde\beta} |\sigma_{X^*}^2)$, $Z_1' \sim N(0,|{b}/{\widetilde\beta} |\sigma_{X^*}^2)$, and $Z_2 \sim N(0,|b\beta|\sigma_{X^*}^2 )$ be independent random variables.
If $b\beta>0$ then define
${\widetilde X}^* = X^* +Z_1$,  
$ \eta= \widetilde \eta+Z'_1$, and 
$\widetilde \eps = \eps +Z_2$
otherwise define
$ X^* = {\widetilde X}^* +Z_1$, 
$ \widetilde \eta=  \eta+Z'_1$, and 
$ \eps = \widetilde \eps +Z_2$.

Then $\varphi_{\widetilde{X}^*}(u)=\varphi_{ X^*}(u)-\alpha_1u^2=-(\frac{1}{2}\sigma_{X^*}^2+\alpha_1)u^2$,
$\varphi_{\widetilde \eta}(s_1)=\varphi_{ \eta}(s_1)+\alpha_1s_1^2$, and $\varphi_{ \widetilde\eps}(s_2)=\varphi_{ \eps}(s_2)+\alpha_2s_2^2$, 
where $\alpha_1=\frac{1}{2}\sigma_{X^*}^2{b}/{\widetilde\beta} $ and
$\alpha_2=-\frac{1}{2} b\beta\sigma_{X^*}^2$.
Now substituting into \eqref{eq:lcfY2} we obtain,
\begin{align*}
\varphi_{X, Y}(s_1,s_2 )
&=\varphi_{X^*}(s_1+ s_2\beta) 
+ \varphi_{\eta}(s_1)
+\varphi_{ \eps}(s_2)\\
&=-\frac{1}{2}\sigma_{X^*}^2(s_1+ s_2\beta)^2 
+ \left(\varphi_{\widetilde \eta}(s_1)-\frac{1}{2}\frac{b\sigma_{X^*}^2}{\widetilde\beta}s_1^2\right)
+\left(\varphi_{\widetilde \eps}(s_2)+\frac{1}{2} b\beta\sigma_{X^*}^2s_2^2\right)\\
&=-\frac{1}{2}\sigma_{X^*}^2s_1^2-\sigma_{X^*}^2\beta s_1s_2 -\frac{1}{2}\sigma_{X^*}^2 s_2^2\beta^2 
-\frac{1}{2}\frac{b\sigma_{X^*}^2}{\widetilde\beta}s_1^2
+\frac{1}{2} b\beta\sigma_{X^*}^2s_2^2+\varphi_{\widetilde \eta}(s_1)+\varphi_{\widetilde \eps}(s_2)\\
&=-\frac{1}{2}\sigma_{X^*}^2\frac{\beta}{\widetilde \beta}s_1^2
-\sigma_{X^*}^2\beta s_1s_2 
-\frac{1}{2}\beta \widetilde \beta\sigma_{X^*}^2s_2^2
+\varphi_{\widetilde \eta}(s_1)+\varphi_{\widetilde \eps}(s_2)\\
&=-\frac{1}{2}\sigma_{X^*}^2\frac{\beta}{\widetilde \beta}(s_1
+ \widetilde \beta s_2)^2
+\varphi_{\widetilde \eta}(s_1)+\varphi_{\widetilde \eps}(s_2)\\
&=\varphi_{{\widetilde X}^*}(s_1+ s_2\widetilde \beta)
+ \varphi_{\widetilde \eta}(s_1)
+\varphi_{ \widetilde\eps}(s_2).
\end{align*}

\subsection{Proof of Theorem \protect\ref{th:asym}} \label{ap:asym}

The estimator in \eqref{eq:IGMM} minimizes the integral of a GMM norm. 
The proof of root-$n$ asymptotic normality is basically the same as the proof for any extremum estimator
\citep[see, e.g.,][]{newey1994large}.
See \cite{carrasco2000generalization} for a proof in a general setting with a continuum of moment conditions and with higher level assumptions.

Define
\begin{align*}
\bm H(\bm b,v) &=\left( \cov (X_{1},X_{2})
+ \left. \frac{ {\partial^{2}	\varphi_{\bm  X,Y}({\bm  s})}}{ \partial s_{1} \partial s_{2}} 
\right|_{\bm s =  (\bm b,-1) v},\ldots, \cov (X_{p-1},X_{p})
+ \left. \frac{ {\partial^{2}	\varphi_{\bm  X,Y}({\bm  s})}}{ \partial s_{p-1} \partial s_{p}} 
\right|_{\bm s =  (\bm b,-1) v},
\right. \\
&\quad \quad \left.
\cov (X_{1},Y)
+ \left. \frac{ {\partial^{2}	\varphi_{\bm  X,Y}({\bm  s})}}{ \partial s_{1} \partial s_{p+1}} 
\right|_{\bm s =  (\bm b,-1) v},\ldots,
\cov (X_{p},Y)
+ \left. \frac{ {\partial^{2}	\varphi_{\bm  X,Y}({\bm  s})}}{ \partial s_{p} \partial s_{p+1}} 
\right|_{\bm s =  (\bm b,-1) v} \right)', \\ 
\bm H_n(\bm b,v) &=\left(\widehat \cov (X_{1},X_{2})
+ \left. \frac{\widehat {\partial^{2}	\varphi_{\bm  X,Y}({\bm  s})}}{ \partial s_{1} \partial s_{2}} 
\right|_{\bm s =  (\bm b,-1) v},\ldots,\widehat \cov (X_{p-1},X_{p})
+ \left. \frac{\widehat {\partial^{2}	\varphi_{\bm  X,Y}({\bm  s})}}{ \partial s_{p-1} \partial s_{p}} 
\right|_{\bm s =  (\bm b,-1) v},
\right. \\
&\quad \quad \left.
\widehat \cov (X_{1},Y)
+ \left. \frac{\widehat {\partial^{2}	\varphi_{\bm  X,Y}({\bm  s})}}{ \partial s_{1} \partial s_{p+1}} 
\right|_{\bm s =  (\bm b,-1) v},\ldots,
\widehat \cov (X_{p},Y)
+ \left. \frac{\widehat {\partial^{2}	\varphi_{\bm  X,Y}({\bm  s})}}{ \partial s_{p} \partial s_{p+1}} 
\right|_{\bm s =  (\bm b,-1) v} \right)',
\end{align*}
where $\bm b\in \mathcal{B} \subset \mathbb{R}^{p}$,
$\bm H(\bm b,v)'$ is the conjugate transpose of $H(\bm b,v)\in \mathbb{C}^{p(p+1)/2\times 1}$, which  is finite for all $\bm b \in \mathcal{B}$ and $v$ almost everywhere by Assumptions \ref{as:reg}(ii)-(iii),
$\bm H_n(\bm b,v)'$ is the conjugate transpose of $\bm H_n(\bm b,v)\in \mathbb{C}^{p(p+1)/2\times 1}$, and 
\begin{align*}
\widehat \cov(X_{k_1},X_{k_2})
&=\frac{1}{n}\sum_{i=1}^nX_{ik_1}X_{ik_2}-\left(\frac{1}{n}\sum_{i=1}^nX_{ik_1}\right)\left(\frac{1}{n}\sum_{i=1}^nX_{ik_2}\right),\\
\widehat \cov(X_{k},Y)
&=\frac{1}{n}\sum_{i=1}^nX_{ik}Y_{i}-\left(\frac{1}{n}\sum_{i=1}^nX_{ik}\right)\left(\frac{1}{n}\sum_{i=1}^nY_i\right),\\
\left. \frac{\widehat {\partial^{2}	\varphi_{\bm  X,Y}({\bm  s})}}{ \partial s_{k_1} \partial s_{k_2}} \right|_{\bm s=(\bm b,-1)v}
&=\frac{\left(\frac{1}{n} \sum_{i=1}^n X_{ik_1} e^{\textup{i}v(\bm b'\bm X_i -Y_i)} \right)
	\left(\frac{1}{n} \sum_{i=1}^n X_{ik_2} e^{\textup{i} v(\bm b'\bm X_i -Y_i)}  \right)} 
{ \left(\frac{1}{n} \sum_{i=1}^n e^{\textup{i} v(\bm b'\bm X_i -Y_i)} \right)^2} 
- \frac{ \frac{1}{n} \sum_{i=1}^nX_{ik_1}X_{ik_2} e^{\textup{i} v(\bm b'\bm X_i -Y_i)}   }
{ \frac{1}{n} \sum_{i=1}^n e^{\textup{i} v(\bm b'\bm X_i -Y_i)} }, \\
\left. \frac{\widehat {\partial^{2}	\varphi_{\bm  X,Y}({\bm  s})}}{ \partial s_{k} \partial s_{p+1}} \right|_{\bm s=(\bm b,-1)v}
&=\frac{\left(\frac{1}{n} \sum_{i=1}^n X_{ik} e^{\textup{i}v(\bm b'\bm X_i -Y_i)} \right)
	\left(\frac{1}{n} \sum_{i=1}^n Y_{i} e^{\textup{i} v(\bm b'\bm X_i -Y_i)}  \right)} 
{ \left(\frac{1}{n} \sum_{i=1}^n e^{\textup{i} v(\bm b'\bm X_i -Y_i)} \right)^2} 
- \frac{ \frac{1}{n} \sum_{i=1}^nX_{ik}Y_{i} e^{\textup{i} v(\bm b'\bm X_i -Y_i)}   }
{ \frac{1}{n} \sum_{i=1}^n e^{\textup{i} v(\bm b'\bm X_i -Y_i)} }.
\end{align*}
Define
\begin{align*}
Q\left( \bm b\right) &=-\int_{-\infty}^{\infty}\bm H(\bm  b,v)'\bm W(v) \bm H( \bm b,v) \pi(v) dv,\\
Q_n \left( \bm b\right) &=- \int_{-\infty}^{\infty}\bm H_n( \bm b,v)'\bm W_n(v)\bm H_n( \bm b,v) \pi(v) dv, 
\end{align*}
where  
$\bm W_n(v) \stackrel{p}{\rightarrow}\bm W(v)$ almost everywhere for some positive definite matrix $\bm W(v)$ by Assumption \ref{as:reg}(iv) and $\pi(v)>0$ is a probability density function. 
Then the IGMM estimator is,
\begin{align*}
\widehat {\bm \beta}  = \argmax_{\bm b \in \mathcal{B}} Q_n \left( \bm b\right) .
\end{align*}
We show that under Assumption \ref{as:nongaus}, $Q$ is maximized if and only if 
$\bm b=\bm \beta$ or equivalently that $\bm H(\bm b,v)=0$  if and only if 
$\bm b=\bm \beta$. The log characteristic function of  $(\bm X,Y)$  is
\begin{align*}
\varphi_{\bm X, Y}(\bm s )
&=\varphi_{\bm X^*}((\bm I_{p}  , \bm \beta)\bm s) 
+\varphi_{ \bm \eta, \eps}(\bm s),  \qquad \bm s \in \mathbb{R}^{p+1},
\end{align*}
where $\varphi_{ \bm \eta, \eps}(\bm s) =\varphi_{ \eps}(s_{p+1}) + \sum_{k=1}^p\varphi_{\eta_k}(s_k)$ when Assumption \ref{as:eivdep}(i) holds and $\varphi_{ \bm \eta, \eps}(\bm s)=-\frac{1}{2} \bm s'\bm \Sigma \bm s$ when Assumption \ref{as:eivdep}(ii) holds.
The second derivatives evaluated at $\bm s= (\bm b,-1)  v$ less the derivatives evaluated at $\bm s=\bm 0$ are,
\begin{align*}
\cov (X_{k_1},Y)
+  \left.\frac{{\partial^{2}	\varphi_{\bm  X,Y}({\bm  s})}}{ \partial s_{k_1} \partial s_{p+1}} \right|_{\bm s= (\bm b,-1)  v}
&=\sum_{k=1}^p \beta_{k}  \left(\cov \left(X^*_{k_1}, X^*_{k}\right)
+  \left. \frac{\partial^2 \varphi_{\bm X^*}(\bm u)}{\partial u_{k_1}\partial u_{k}}   \right|_{\bm u =( \bm b-\bm \beta)	v}
\right), \quad 1\leq k_1  \leq p,\\
\cov (X_{k_1},X_{k_2})
+ \left. \frac{{\partial^{2}	\varphi_{\bm  X,Y}({\bm  s})}}{ \partial s_{k_1} \partial s_{k_2}}  \right|_{\bm s= (\bm b,-1)  v}
&=\cov \left(X^*_{k_1}, X^*_{k_2}\right)
+  \left. \frac{\partial^2 \varphi_{\bm X^*}(\bm u)}{\partial u_{k_1}\partial u_{k_2}}   \right|_{\bm u =( \bm b-\bm \beta)	v}
, \quad 1\leq k_1<k_2  \leq p,
\end{align*}
where all derivatives are finite because of Assumptions \ref{as:reg}(ii)-(iii).
The above expressions are equal to zero when $\bm b =\bm \beta$ so $Q$ is maximized at $\bm b=\bm \beta$ because $Q(\bm \beta)=0$.
Now assume that $\widetilde{\bm \beta}\neq \bm \beta$ and  $Q(\widetilde{\bm \beta})=0$. Then, 
\begin{align*}
0
&=\sum_{k=1}^p \beta_{k}  \left(\cov \left(X^*_{k_1}, X^*_{k}\right)
+  \left. \frac{\partial^2 \varphi_{\bm X^*}(\bm u)}{\partial u_{k_1}\partial u_{k}}   \right|_{\bm u =( \widetilde{\bm \beta}-\bm \beta)	v}
\right), \quad 1\leq k_1  \leq p,\\
0
&=\cov \left(X^*_{k_1}, X^*_{k_2}\right)
+  \left. \frac{\partial^2 \varphi_{\bm X^*}(\bm u)}{\partial u_{k_1}\partial u_{k_2}}   \right|_{\bm u =( \widetilde{\bm \beta}-\bm \beta)	v}
, \quad 1\leq k_1<k_2  \leq p,
\end{align*}
and so 
\begin{align*}
\left. \frac{\partial^2 \varphi_{\bm X^*}(\bm u)}{\partial u_{k_1}\partial u_{k_2}}   \right|_{\bm u =(\widetilde{\bm \beta}-\bm \beta)	v}
&=-\cov \left(X^*_{k_1}, X^*_{k_2}\right), \quad 1\leq k_1 \leq k_2  \leq p.
\end{align*}
We can now twice apply the fundamental theorem of calculus, because all the functions are continuous by Assumptions \ref{as:reg}(ii)-(iii), to obtain
$
\varphi_{\bm X^*}((\widetilde{\bm \beta}-\bm \beta )v)	=P_2(v)
$. 
Hence, by \cite{marcinkiewicz1939propriete}, $(\widetilde{\bm \beta}-\bm \beta )'\bm X^*$ is normal.
This is a contradiction by Assumption \ref{as:nongaus}. Hence $Q$ is uniquely maximized at $\bm b=\bm \beta$.

By Assumptions \ref{as:iid} and \ref{as:reg}(i)-(iii), we have 
$\frac{\partial \bm H_n}{\partial \bm b'}( \bm b,v)\stackrel{p}{\rightarrow} \frac{\partial \bm H}{\partial \bm b'}( \bm b,v) $ and $\bm H_n( \bm b,v) \stackrel{p}{\rightarrow} \bm H( \bm b,v)$ almost everywhere and by the standard argument  $\widehat {\bm \beta} \stackrel{p}{\rightarrow} \bm \beta$. We now 
use Assumption \ref{as:reg}, consistency of $\widehat {\bm \beta}$, and identification of $\bm \beta$  to show asymptotic normality. 

Let $n$ be large enough. 
The first order conditions of \eqref{eq:IGMM} are,
\begin{align*}
\bm 0 &=\int_{-\infty}^{\infty}\frac{\partial \bm H_n}{\partial \bm b}( \widehat {\bm \beta},v)\bm W_n(v)\bm H_n( \widehat {\bm \beta},v) \pi(v) dv\\
&=\int_{-\infty}^{\infty}\frac{\partial \bm H_n}{\partial \bm b}( \widehat {\bm \beta},v)\bm W_n(v)\left(\bm H_n(  {\bm \beta},v)+ \frac{\partial \bm H_n}{\partial \bm b'}( \widetilde{\bm \beta},v)(\widehat {\bm \beta}-\bm \beta)\right) \pi(v) dv,
\end{align*}
where 
the first equality follows by Assumption \ref{as:reg}(v), and the second equality by the mean value theorem that uses Assumptions \ref{as:reg}(ii)-(iii) ($\widetilde{\bm \beta}$ is on the line segment joining $\widehat {\bm \beta}$ and $\bm \beta$).
Rearranging and multiplying by $\sqrt{n}$, 
{\small
	\begin{align*}
	\sqrt{n}(\widehat {\bm \beta}-\bm \beta)&=-\left( \int_{-\infty}^{\infty}\frac{\partial \bm H_n}{\partial \bm b}( \widehat {\bm \beta},v)\bm W_n(v)\frac{\partial \bm H_n}{\partial \bm b'}( \widetilde{\bm \beta},v) \pi(v) dv\right)^{-1}\left(\int_{-\infty}^{\infty}\frac{\partial \bm H_n}{\partial \bm b}( \widehat {\bm \beta},v)\bm W_n(v) \sqrt{n} \bm H_n(  {\bm \beta},v)\pi(v)dv\right),
	\end{align*}
}where the matrix is invertible by Assumption \ref{as:reg}(vi).
Now we linearize  $\bm H_n(  {\bm \beta},v)$ to obtain,
{\small
	\begin{align*}
	&\sqrt{n}\bm H_n(  {\bm \beta},v)= \Big( \cdots \\ 
	&\sqrt{n} \frac{1}{n}\sum_{i=1}^n \left\{a_{k_2}(v) \left(X_{ik_1}e^{\textup{i}v(\bm \beta'\bm X_i -Y_i)} - E[X_{k_1}e^{\textup{i}v(\bm \beta'\bm X -Y)}]\right) +a_{k_1}(v) \left(X_{ik_2}e^{\textup{i}v(\bm \beta'\bm X_i -Y_i)} - E[X_{k_2}e^{\textup{i}v(\bm \beta'\bm X -Y)}]\right)\right.\\
	& \left.  +  a_{0}(v) \left(X_{ik_1}X_{ik_2}e^{\textup{i}v(\bm \beta'\bm X_i -Y_i)} - E[X_{k_1}X_{k_2}e^{\textup{i}v(\bm \beta'\bm X -Y)}]\right) +a_{0k_1k_2}(v) \left(e^{\textup{i}v(\bm \beta'\bm X_i -Y_i)} - E[e^{\textup{i}v(\bm \beta'\bm X -Y)}]\right)\right\}+o_p(1),\\
	&\qquad \qquad \qquad \quad \cdots \Big),
	\end{align*}
}where 
\begin{align*}
a_{k_1}(v)&=\frac{E[X_{k_1}e^{\textup{i}v(\bm \beta'\bm X -Y)}]}{(E[e^{\textup{i}v(\bm \beta'\bm X -Y)}])^2} ,
\qquad a_{k_2}(v)=\frac{E[X_{k_2}e^{\textup{i}v(\bm \beta'\bm X -Y)}]}{(E[e^{\textup{i}v(\bm \beta'\bm X -Y)}])^2} ,
\qquad a_{0}(v)=\frac{1}{(E[e^{\textup{i}v(\bm \beta'\bm X -Y)}])^2} ,\\
a_{0k_1k_2}(v)&=\frac{E[X_{k_1}X_{k_2}e^{\textup{i}v(\bm \beta'\bm X -Y)}]E[e^{\textup{i}v(\bm \beta'\bm X -Y)}]
	-2E[X_{k_1}e^{\textup{i}v(\bm \beta'\bm X -Y)}]E[X_{k_2}e^{\textup{i}v(\bm \beta'\bm X -Y)}]
}{(E[e^{\textup{i}v(\bm \beta'\bm X -Y)}])^2} ,
\end{align*}
are finite by Assumptions \ref{as:reg}(ii)-(iii).
Using Assumptions \ref{as:reg}(v)-(vi) to apply Slutsky's Theorem and Fatou's Lemma, and also Assumption \ref{as:iid} to apply the Lindeberg-Feller central limit theorem,
\begin{align*}
\sqrt{n}(\widehat {\bm \beta}-\bm \beta)
&\stackrel{d}{\rightarrow}N\left(\bm 0,\bm S \left(\int_{-\infty}^{\infty} \int_{-\infty}^{\infty}
\frac{\partial \bm H}{\partial \bm b}(  {\bm \beta},v)\bm W(v)\bm \Omega(v,w)\bm W(w) \frac{\partial \bm H}{\partial \bm b'}(  {\bm \beta},w)\pi(v)\pi(w)dvdw\right)\bm S'
\right),
\end{align*}
where
$
\bm S = \left( \int_{-\infty}^{\infty}\frac{\partial \bm H}{\partial \bm b}( {\bm \beta},v)\bm W(v)\frac{\partial \bm H}{\partial \bm b'}(  {\bm \beta},v) \pi(v) dv\right)^{-1}
$
and the terms in $\bm \Omega(v,w)$ are of the form,
{\small
\begin{align}	
&\cov\left(a_{k_2}(v) X_{k_1}e^{\textup{i}v(\bm \beta'\bm X -Y)} +a_{k_1}(v) X_{k_2}e^{\textup{i}v(\bm \beta'\bm X -Y)}
+  a_{0}(v) X_{k_1}X_{k_2}e^{\textup{i}v(\bm \beta'\bm X -Y)}  
+a_{0k_1k_2}(v) e^{\textup{i}v(\bm \beta'\bm X -Y)},\right.  \nonumber\\
&   \left. a_{k_4}(w) X_{k_3}e^{\textup{i}w(\bm \beta'\bm X -Y)} 
+  a_{k_3}(w) X_{k_4}e^{\textup{i}w(\bm \beta'\bm X -Y)}
+ a_{0}(w) X_{k_3}X_{k_4}e^{\textup{i}w(\bm \beta'\bm X -Y)} +a_{0k_3k_4}(w)e^{\textup{i}  w(\bm \beta'\bm X -Y)} \right). \label{eq:Omegaterms}
\end{align}
}
\end{document}